\documentclass[letterpaper]{article} % DO NOT CHANGE THIS
\usepackage[draft]{aaai2026}  % DO NOT CHANGE THIS
\usepackage{times}  % DO NOT CHANGE THIS
\usepackage{helvet}  % DO NOT CHANGE THIS
\usepackage{courier}  % DO NOT CHANGE THIS
\usepackage[hyphens]{url}  % DO NOT CHANGE THIS
\usepackage{graphicx} % DO NOT CHANGE THIS
\usepackage{natbib}  % DO NOT CHANGE THIS AND DO NOT ADD ANY OPTIONS TO IT
\usepackage{caption} % DO NOT CHANGE THIS AND DO NOT ADD ANY OPTIONS TO IT
\usepackage{algorithmic}
\usepackage{newfloat}
\usepackage{listings}
\DeclareCaptionStyle{ruled}{labelfont=normalfont,labelsep=colon,strut=off} % DO NOT CHANGE THIS
\title{SenseCrypt: Sensitivity-guided Selective Homomorphic Encryption for Cross-Device Federated Learning}

\author{
   Borui Li \textsuperscript{\rm 1},
   Li Yan \textsuperscript{\rm 1},
   Junhao Han \textsuperscript{\rm 1},
   Jianmin Liu \textsuperscript{\rm 1},
   Lei Yu \textsuperscript{\rm 2}
}
\affiliations{
    \textsuperscript{\rm 1} School of Cyber Science and Engineering\\
      Xi'an Jiaotong University\\
      Shaanxi, China 710049 \\
    \textsuperscript{\rm 2} Department of Computer Science \\
      Rensselaer Polytechnic Institute \\
      Troy, NY USA 12180 \\

    boruili@stu.xjtu.edu.cn
}

\usepackage{bibentry}
\usepackage{booktabs}
\usepackage{multirow}
\usepackage{subcaption}
\usepackage[para]{threeparttable}
\usepackage{graphbox}
\usepackage{amsmath}
\usepackage[ruled, vlined, linesnumbered]{algorithm2e}
\usepackage{tabularray}
\usepackage{array} 
\usepackage{amssymb}
\usepackage{tikz}
\usepackage{enumitem}
\usepackage{amsfonts}

\SetFuncSty{textsf}
\SetKwProg{Fn}{}{:}{}

\newcommand{\squishlist}{
\begin{list}{$\bullet$}
  { \setlength{\itemsep}{0pt}
     \setlength{\parsep}{0pt}
     \setlength{\topsep}{0pt}
     \setlength{\partopsep}{0pt}
     \setlength{\leftmargin}{0em}
     \setlength{\labelwidth}{0em}
     \setlength{\labelsep}{0.2em} } }

\newcommand{\squishlisttwo}{
\begin{list}{$\bullet$}
  { \setlength{\itemsep}{0pt}
     \setlength{\parsep}{0pt}
    \setlength{\topsep}{0pt}
    \setlength{\partopsep}{0pt}
    \setlength{\leftmargin}{2em}
    \setlength{\labelwidth}{1.5em}
    \setlength{\labelsep}{0.5em} } }

\newcommand{\squishend}{
  \end{list}  }

\begin{document}

\maketitle

\begin{abstract}
Homomorphic Encryption (HE) prevails in securing Federated Learning (FL), but suffers from high overhead and adaptation cost. 
Selective HE methods, which partially encrypt model parameters by a global mask, are expected to protect privacy with reduced overhead and easy adaptation. 
However, in cross-device scenarios with heterogeneous data and system capabilities, traditional Selective HE methods deteriorate client straggling,
and suffer from degraded HE overhead reduction performance. %, while the model parameter sensitivity is effective for clients' data distribution similarity measurement. %, hence not applicable for cross-device FL.
Accordingly, we propose \emph{SenseCrypt}, a \underline{\textbf{Sen}}sitivity-guided \underline{\textbf{se}}lective Homomorphic En\underline{\textbf{Crypt}}ion framework, to adaptively balance security and HE overhead per cross-device FL client.
Given the observation that model parameter sensitivity is effective for measuring clients' data distribution similarity, we first design a privacy-preserving method to respectively cluster the clients with similar data distributions.
Then, we develop a scoring mechanism to deduce the straggler-free ratio of model parameters that can be encrypted by each client per cluster.
Finally, for each client, we formulate and solve a multi-objective model parameter selection optimization problem, which minimizes HE overhead while maximizing model security without causing straggling.
Experiments demonstrate that \emph{SenseCrypt} ensures security against the state-of-the-art inversion attacks, while achieving normal model accuracy as on IID data, and reducing training time by 58.4\%$\sim$88.7\% as compared to traditional HE methods.
\end{abstract}

\section{Introduction} \label{sec:intro}

Federated Learning (FL) is increasingly popular due to its ability to enable collaborative Machine Learning (ML) model training 
without exposing privacy-sensitive user data \cite{DBLP:journals/spm/LiSTS20,kairouz2021advances}.
However, multiple studies have shown that the model parameters transmitted during FL training can be exploited via \emph{inversion attacks} to reconstruct users' private data \cite{DBLP:conf/ccs/HitajAP17,dlg,geiping2020inverting}.
To defend against such attacks, Homomorphic Encryption (HE) has attracted much attention due to its strong privacy guarantee while enabling mathematical operation of encrypted model parameters \cite{zhang2020batchcrypt,roth2022nvidia,jiang2021flashe,8761267,hanyan,chen2021secureboost,xu2021efficient,zheng2022aggregation,jin2023fedmlhe,hu2024maskcrypt}.

Nevertheless, applying HE per parameter (\emph{Plain} HE) is computationally and communicatively expensive \cite{zhang2020batchcrypt,jin2023fedmlhe}. To mitigate this, various optimization strategies have been proposed. For instance, \textbf{Lightweight HE} reduces cryptographic complexity at the cost of weaker security \cite{jiang2021flashe,8761267}, while \textbf{Batch HE} encrypts multiple parameters together but can introduce accuracy loss from quantization or require non-standard framework adaptations \cite{zhang2020batchcrypt,hanyan,chen2021secureboost,xu2021efficient,zheng2022aggregation}. In contrast, \textbf{Selective HE} draws inspiration from model pruning research \cite{lecun1989optimal}, which demonstrates that many model parameters are non-essential. This approach encrypts only the most critical parameters identified via a shared \emph{selective encryption mask}. A key advantage is its high compatibility with existing FL frameworks, as it avoids altering the core encryption or aggregation algorithms \cite{jin2023fedmlhe,hu2024maskcrypt}.
However, existing Selective HE methods fail to balance security and efficiency in cross-device FL with Non-IID 
data and heterogeneous device capabilities.
Our experimental studies in 
Appendix Sections C and D 
demonstrate that 1) the uniform encryption budgets (defined as the upper bound quantity for encrypting model parameters) used in these methods may exacerbate the straggler problem on clients with heterogeneous device capabilities (i.e., \emph{system heterogeneity}), and 2) their unioned encryption masks may incur prohibitive overhead when the clients' data distributions diverge (i.e., \emph{statistical heterogeneity}).

Driven by these limitations,
we aim to adaptively balance security and HE overhead for each client in cross-device FL.
To achieve this, we need to address the following challenges:
\begin{itemize}[
fullwidth,
topsep=0pt,
      itemindent=1em,
]

    \item 
    \textbf{How to avoid the negative impact of statistical and system heterogeneity on Selective HE?}
    Multiple solutions have been proposed to deal with statistical heterogeneity issues in FL via privacy-preserving data similarity measurement and clustering of clients with IID data \cite{cho2022flame,liu2021pfa}.
    In the meantime, to mitigate the straggler problem, 
    several methods have been proposed to screen clients by their training performance \cite{bonawitz2019towards,chai2020tifl,zhou2023reinforcement,jiang2023heterogeneity}.
    However, most of them rely on additional components (e.g., sparsity of Convolutional Neural Networks,
    client scheduler)
    to measure data similarity or coordinate FL training, which creates extra burden for FL framework adaptation.
    We observe that in addition to weighing model parameter criticality in defending against inversion attacks, 
    model parameter sensitivity can reflect data distribution as well (Appendix Section E).
    Given this fact, a non-intrusive while privacy-preserving method for guiding the selective HE of cross-device FL clients is expected. %but non-trivial.

    \item 
    \textbf{How to adaptively balance security and HE overhead for each client?}
    Encrypting too many model parameters on clients with low device capability significantly degrades FL performance, while encrypting insufficient parameters increases data leakage risk.
    However, without a sound measurement of the clients' device capability and security impact caused by selective HE, it is challenging to adaptively tailor the straggler-free encryption budget per client,
    while ensuring sufficient model security.

\end{itemize}

Accordingly, we propose \emph{SenseCrypt}, a \underline{\textbf{Sen}}sitivity-guided \underline{\textbf{se}}lective Homomorphic En\underline{\textbf{Crypt}}ion framework, to adaptively achieve the optimal balance between model security and HE overhead for cross-device FL clients.
Specifically, by exploiting model parameter sensitivity to represent data distribution and similarity measurement, we first design a privacy-preserving method to cluster the clients with IID data into their respective groups.
Then, based on the bandwidth and device processing speed of the clients in FL,
we establish a scoring mechanism to deduce the straggler-free encryption budget for each client per cluster. %, which won't deteriorate the straggler problem within the cluster.
Finally, based on the sensitivity assessment result and encryption budget, we let each client  formulate and solve a multi-objective model parameter selection optimization problem that aims to minimize HE overhead while maximizing model security without causing straggling.
\section{Preliminaries and Motivations} 
\label{sec:why}

\subsection{Definition of Model Parameter Sensitivity} \label{sec:why:def}

Suppose $\mathbf{W}$ represents the model parameters of a neural network. 
We denote the model loss function as $\mathcal{L}(\mathbf{W})$, and the gradients of the loss with respect to $\mathbf{W}$ as $\nabla_{\mathbf{W}}\mathcal{L}(\mathbf{W})$.
Given a subset of the model parameters $\mathbf{w} \in \mathbf{W}$, 
the sensitivity of $\mathbf{w}$ is denoted as $\mathbf{\Gamma}(\mathbf{w}) \in \mathbb{R}^{|\mathbf{w}|}$, and the model parameters with $\mathbf{w}$ zeroed-out is denoted as $\mathbf{W}_{-\mathbf{w}}$.
The sensitivity of $\mathbf{w}$ is defined as the change in the loss function after zeroing $\mathbf{w}$ \cite{lecun1989optimal}:

\begin{equation} 
\label{equ:sen}
    \mathbf{\Gamma}(\mathbf{w}) = \lvert \mathcal{L}(\mathbf{W}) - \mathcal{L}(\mathbf{W}_{-\mathbf{w}}) \rvert.
\end{equation}

From Equation (\ref{equ:sen}), % we know that 
the larger absolute loss change caused by the removal of $\mathbf{w}$, the more contribution $\mathbf{w}$ can make to the loss, and the more sensitive $\mathbf{w}$ is.
Thus, $\mathbf{\Gamma}(\mathbf{w})$ also corresponds to the privacy risk on gradients that may be exposed via observing $\mathbf{w}$.
Considering that it is computationally infeasible to calculate the sensitivity of any arbitrary subset of model parameters by forward-passing the model every time, we adopt the first-order Taylor expansion of $\mathcal{L}(\cdot)$ with respect to $\mathbf{w}$ at $\mathbf{W}$ to approximate $\mathbf{\Gamma}(\mathbf{w})$ as in \cite{DBLP:conf/emnlp/XuKM22,DBLP:conf/iclr/LiangJZH0GCZ22}:

\begin{equation} \label{equ:taylor}
    \mathbf{\Gamma}(\mathbf{w}) \approx \lvert \mathbf{w}^\top \nabla_{\mathbf{W}}\mathcal{L}(\mathbf{W}) \rvert.
\end{equation}

Since $\mathbf{w}$ and $\nabla_{\mathbf{W}}\mathcal{L}(\mathbf{W})$ can be easily obtained per client during FL training, the calculation overhead of $\mathbf{\Gamma}(\mathbf{w})$ is trivial as compared to the additional components used by the previous statistical and system heterogeneity solutions \cite{bonawitz2019towards,chai2020tifl,zhou2023reinforcement,jiang2023heterogeneity}.

\subsection{Privacy Leakage Evaluation}
\label{mutual_information}
We employ Mutual Information (MI) to quantify the privacy leakage caused by the selective encryption of model parameters. %as in \cite{mai2024split,wang2021privacy,cuff2016differential}.
Given $\mathbf{W}$ and $\mathbf{W}_{-\mathbf{w}}$, their mutual information is defined as the expected value of the point-wise mutual information over their joint distribution:

\begin{equation}\label{eq:MI}
I(\mathbf{W}; \mathbf{W}_{-\mathbf{w}}) = 
\sum_{y \in \mathbf{W}_{-\mathbf{w}} } \sum_{x \in \mathbf{W}} p(x, y) \log_2 \frac{p(x, y)}{p(x) p(y)}
\end{equation}
where $p(x,y)$ represents the joint probability distribution, $p(x)$ and $p(y)$ represent the marginal distributions of $\mathbf{W}$ and $\mathbf{W}_{-\mathbf{w}}$ respectively. 
Considering that we aim to selectively encrypt the most privacy-sensitive model parameters $\mathbf{w}$, thus $I(\mathbf{W}; \mathbf{W}_{-\mathbf{w}})$ reflects the amount of privacy that can be leaked via the unencrypted ones 
(i.e., $\mathbf{W}_{-\mathbf{w}}$). 
The larger $I(\mathbf{W}; \mathbf{W}_{-\mathbf{w}})$ is, the higher privacy leakage risk that the selective encryption of $\mathbf{w}$ will result in.

\subsection{Threat Model} 
\label{sec:secmodel}

Following existing Selective HE methods \cite{hu2024maskcrypt,jin2023fedmlhe}, we assume an \emph{honest-but-curious} server that follows the FL protocol, yet attempts to infer sensitive information through inversion attacks \cite{DBLP:conf/ccs/HitajAP17,dlg,geiping2020inverting}. 
We assume that the clients will use their public keys to selectively encrypt model parameters, and will not share their private keys to the server or the other clients as in
\cite{zhang2020batchcrypt,jin2023fedmlhe}.
We also assume that
attackers
can only launch inversion attacks via unencrypted model parameters. %, which means it cannot interfere with FL algorithms, modify model architecture, or send malicious global model parameters to better assist in its attacks.
The protection against other forms of malicious activities 
is not the focus of this work, and we refer to existing methods for protection \cite{zheng2022aggregation,QueyrutSF23}.
To extend our threat model to defend against client collusion attacks, we introduce a dual-server setting in Appendix I. This setting decouples the key distribution and decryption operations, while the rest of the algorithmic procedure remains unchanged. The pseudocode is provided in Algorithm 2 in the Appendix.
\section{System Design of \emph{SenseCrypt}}\label{sec:sys}

\begin{figure}
\centering
    \includegraphics[width=.9\linewidth]{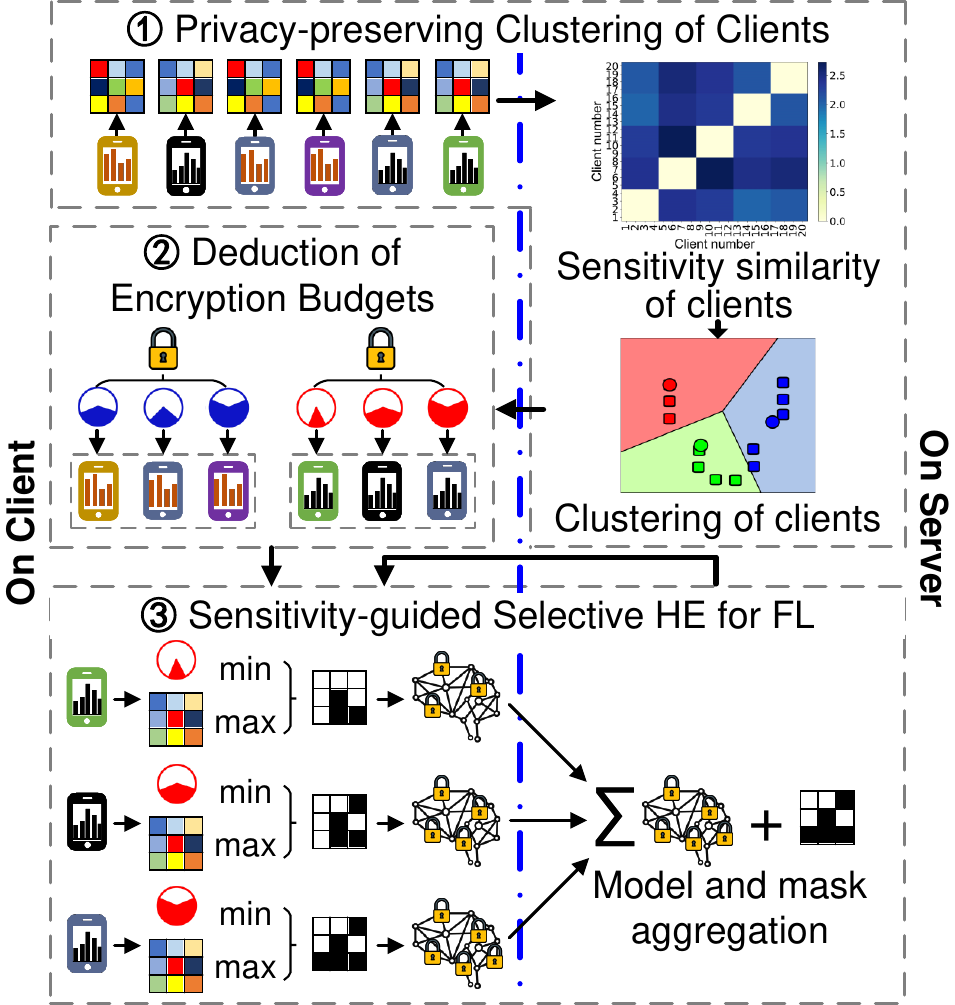}
    \caption{Framework of \emph{SenseCrypt}.}
    \label{fig:framework}
\end{figure}

\subsection{Privacy-preserving Clustering of Clients 
} \label{sec:sys:clustering}
Given that model parameter sensitivity reflects data distribution (Appendix Section E), 
we aim to exploit it for privacy-preserving
clustering of clients, ensuring that those with IID data are clustered together.
Considering that both gradients and sensitivity vectors can reveal the change of loss values, we let each client only upload its sensitivity vector once during the first FL training iteration.
Thus, clients' sensitivity vectors cannot be observed
(like gradients) for inversion attack as in \cite{DBLP:conf/ccs/HitajAP17,dlg,geiping2020inverting}.

Specifically, during the first FL training iteration, we apply Equation (\ref{equ:taylor}) to obtain the sensitivity vector of each client, and adopt the Euclidean Distance \cite{danielsson1980euclidean} as the metric of similarity between sensitivity vectors for the sake of simplicity and generality.
For the sensitivity vectors of the $i$-th client (denoted as $\mathbf{\Gamma}_i$) and the $j$-th client (denoted as  $\mathbf{\Gamma}_j$), their similarity is measured as: $\text{sim}(\mathbf{\Gamma}_i, \mathbf{\Gamma}_j) = \sqrt{\sum_{k=1}^{N^{\mathbf{w}}} (\gamma_k^i - \gamma_k^j)^2}$,
where $\gamma_k^i$ and $\gamma_k^j$ denote the $k$-th element in $\mathbf{\Gamma}_i$ and $\mathbf{\Gamma}_j$, respectively, and $N^{\mathbf{w}}$ denotes the total number of model parameters.
Considering that the aggregation server has no prior knowledge on client data distribution (e.g., how many IID clusters the clients belong to), we use the Affinity Propagation (AP) method \cite{frey2007clustering} to cluster the clients.
Different from the K-means method \cite{macqueen1967some},
AP automatically selects cluster centers ``exemplars'' by iteratively passing similarity and responsibility messages between data points, eliminating the need for predefined cluster numbers.
Finally, the clients will be clustered into their respective groups, within which the clients' data distribution is IID to each other.

To preemptively mitigate potential inference attacks targeting sensitivity vectors, we introduce an optional Differential Privacy (DP) noise injection mechanism over sensitivity values. 
Appendix Section G empirically demonstrates the plausibility of privacy-preserving client clustering with DP-noised sensitivity vectors and the inherent robustness of AP to noise perturbations.
The workflow of this component is illustrated as  \textcircled{\small{\textbf{1}}} in Figure \ref{fig:framework}.

\subsection{Deduction of Encryption Budgets} \label{sec:encbud_bound}

To avoid the negative impact of system heterogeneity,
the encryption budgets of different clients must be set according to their system capabilities.
Considering that device computation speed and communication bandwidth are the most decisive factors in determining the device's system capability in FL \cite{chai2020tifl,10228970}, we use them to deduce the clients' respective encryption budgets.

Specifically, suppose that in Section \ref{sec:sys:clustering}, the server has measured the bandwidth of each client per cluster (denoted as $r_i$ for client $i$) based on the training delay and the amount of data transmitted during the iteration. %(denoted as $\{r_1, r_2, \ldots, r_{N^c}\}$), where $N^c$ is the number of clients in the cluster.
Meanwhile, from the device specification of each client, the server can obtain client $i$'s CPU clock speed $v_i$ (e.g., 2.4 GHz$\times$4 cores).
Since the client's system capability in FL will always be bottlenecked by its most limited resource, we propose to use the lesser 
of $r_i$ and $v_i$ to deduce each client's encryption budget.
However, $r_i$ and $v_i$ are measured on different scales and are independent of each other. 
Therefore, to score the clients' relative system capabilities, the server first normalizes them into the same scale (denoted as $\overline{r}_i \in [0,1]$ and $\overline{v}_i \in [0,1]$) via the Max Absolute Scaling: $\overline{u}_i = \frac{u_i}{\max\{\lvert u_i \rvert\}^{N^c}}$, where $u_i \in \{ r_i, v_i \}$ is the original value of either $r_i$ or $v_i$, and $\max\{\lvert u_i \rvert\}^{N^c}$ is the maximum absolute value of $r_i$ or $v_i$ over all the $N^c$ clients in the cluster.
Following this normalization, the system capability of client $i$ can be represented as the minimum of $\overline{r}_i$ and $\overline{v}_i$, i.e., $\min \{ \overline{r}_i, \overline{v}_i \}$.
Finally, the server further normalizes each client's system capability, and sends it back to the client as its encryption budget $\alpha_i = \frac{\min \{ \overline{r}_i, \overline{v}_i \}}{\max \{ \lvert \min \{ \overline{r}_i, \overline{v}_i \} \rvert \}^{N^c}}$.
It is obvious that the fastest client's encryption budget will be 1, and $\alpha_i$ corresponds to the ratio of model parameters that the client can timely encrypt without straggling.
The workflow of this component is illustrated as  \textcircled{\small{\textbf{2}}} in Figure \ref{fig:framework}.

\subsection{Sensitivity-guided Selective HE and Model Aggregation} \label{sec:sys:aggr}
Given the calculated model parameter sensitivity and encryption budget of each client,
we elaborate:
1) the selection of model parameters for encryption on each client, which optimally balances model security and HE overhead 
, and
2) model aggregation without the straggler problem. 

\subsubsection{Model Parameter Selection Optimization Problem} \label{sec:sys:cca:opt}

We use a binary mask vector 
$\mathbf{X}_i = [ x_k^i | k = 1, \ldots, N^{\mathbf{w}} ] \in \{0,1\}^{N^{\mathbf{w}}}$
to represent the respective encryption decision of model parameters.
Specifically, %$x_k^i$ can only be 0 or 1.
if $x_k^i$ is 1, it means the model parameter corresponding to $x_k^i$ will be encrypted, or remain unencrypted if otherwise.
On the one hand, we expect to minimize the HE overhead through properly selecting the model parameters for encryption. 
Thus, the first optimization objective can be represented as:

\begin{equation}
\label{equ:obj1}
\min \quad f_1(\mathbf{X}_i) = \sum_{k=1}^{N^{\mathbf{w}}} x_k^i.
\end{equation}
On the other hand, we expect to maximize the protection of the model parameters against privacy risks, i.e., select masks that maximize the sum of sensitivity values corresponding to the masked parameters.
Thus, the other objective can be represented as:

\begin{equation}
\label{equ:sec_target}
\max \quad f_2(\mathbf{X}_i) = \sum_{k=1}^{N^{\mathbf{w}}} x_k^i \gamma_k^i.
\end{equation}

To minimize the straggler problem, the ratio of encrypted model parameters should be lower than the client's encryption budget $\alpha_i$. 
Specifically, %given $\alpha_i$ obtained in Section \ref{sec:encbud_bound},
the straggler-free number of model parameters that the client can encrypt 
is estimated as $\lfloor \alpha_i N^{\mathbf{w}} \rfloor$, where 
$\lfloor \cdot \rfloor$ is the floor function.
Thus, the constraint can be represented as:

\begin{equation}
\label{equ:budget_const}
\sum_{k=1}^{N^{\mathbf{w}}} x_k^i \leqslant \lfloor \alpha_i N^{\mathbf{w}} \rfloor.
\end{equation}

Meanwhile, we must ensure that the total sensitivity measure of protected model parameters 
is sufficiently large to defend against inversion attacks.
On the one hand, we expect the security protection level of the clients generally follows their encryption budgets, as the more model parameters a client can encrypt without straggling, the higher the level of security protection it can achieve \cite{chai2020tifl}.
On the other hand, we don't expect the fast clients to excessively encrypt model parameters for the sake of HE overhead reduction. 
Therefore, we use an exponential function of the
encryption budget to scale the security protection level, which is denoted as $1 - C e^{- B \alpha_i}$, where $C$ is the constant controlling the lower bound of security protection level, and $B$ is the constant controlling the scaling step size of the exponential function.
Compared to linear functions, the exponential function can more effectively prevent clients with excessive computational power from encrypting too many parameters, and renders a more gradual increase in the encryption budget as the client's computational capacity grows.
Thus, the security protection level constraint can be represented as:

\begin{equation}
\label{equ:sec_const}
\frac{1}{\sum_{k=1}^{N^{\mathbf{w}}} \gamma_k^i} \sum_{k=1}^{N^{\mathbf{w}}} x_k^i \gamma_k^i \geqslant 1 - C e^{- B \alpha_i},
\end{equation}
where $\gamma_k^i$
is normalized by the sum of all the model parameters' sensitivity values $\sum_{k=1}^{N^{\mathbf{w}}} \gamma_k^i$ to differentiate the model parameters' relative privacy risk.
A detailed analysis for the selection of $B$ and $C$ is provided in Appendix Section F.
Following the definition in Section \ref{mutual_information}, we add another constraint to limit the privacy leakage caused by selectively encrypting $\mathbf{W}$ into $\mathbf{W}_{-\mathbf{w}}=(1 - \mathbf{X}_i)\odot \mathbf{W}$, where $\odot$ indicates element-wise multiplication:

\begin{equation}
    I(\mathbf{W};\; (1 - \mathbf{X}_i) \odot \mathbf{W}) \leq \eta_{\text{MI}}.
    \label{equ:mi_constraint}
\end{equation}
where $\eta_{\text{MI}}$ is the MI threshold, and can be tuned based on dataset complexity and privacy requirements. 
A lower $\eta_{\text{MI}}$ value enforces stricter privacy guarantee at the cost of higher computational overhead.
A detailed analysis of the relation between MI, attack success rate and encryption overhead is provided in Appendix Section H.

Finally, through combining the objectives and the constraints, the multi-objective model parameter selection optimization problem can be formulated as:

\begin{align}
\min \quad &  f_1(\mathbf{X}_i), \nonumber \\
\max \quad &  f_2(\mathbf{X}_i), \nonumber \\
\textrm{s.t.} 
 \quad & x_k^i \in \{ 0, 1 \}, \\
 & \text{Constraints} (\ref{equ:budget_const}) \;\text{and}\; (\ref{equ:sec_const}) \;\text{and}\; (\ref{equ:mi_constraint}). \nonumber
\end{align}

To solve the Multi-Objective Binary Integer Programming (MOBIP) problem, it is necessary to make the following objective transformation: $F(\mathbf{X}_i) = [\min f_1(\mathbf{X}_i), \max f_2(\mathbf{X}_i)]^\top = \min [f_1(\mathbf{X}_i), -f_2(\mathbf{X}_i)]^\top$.
As $f_l(l = 1,2)$ are separately scaled, we normalize them by 
$\overline{f}_l = \frac{f_l - f_l^{\min}}{f_l^{\max} - f_{l}^{\min}}$, where $f_l^{\min}$ and $f_l^{\max}$ are the maximum and minimum values of $f_l$, respectively.

Then, we scalarize the MOBIP into a single-objective optimization problem: $\min~\beta_1 \overline{f}_1(\mathbf{X}_i) - \beta_2 \overline{f}_2(\mathbf{X}_i)$,
where $\beta_l>0$ $(l = 1, 2)$ are the weights of respective objective functions.  
Thus, the MOBIP problem is transformed to a variant of the Knapsack problem with $N^{\mathbf{w}}$ decision variables. 
Here, we assume equal importance for the objectives by setting $\beta_1 = \beta_2 = 1$, while noting that these weights can be readily customized for other scenarios without altering the overall framework.
By applying Linear Programming (LP) relaxation \cite{cormen2022introduction} on the problem, we can obtain the fractional solutions with $O(N^{\mathbf{w}} \log N^{\mathbf{w}})$ time complexity, which provides a lower bound of the optimal value.
Then, through using algorithms such as rounding or branch and bound, we can further obtain the binary integer solutions to the problem (i.e., the optimal encryption mask vector $\mathbf{X}_i$).
Finally, the client selectively encrypts its model parameters by $\mathbf{X}_i$, and uploads $\mathbf{X}_i$ to the server.
Since the sensitivity of model parameters changes along with the progress of FL training, the sensitivity vector $\mathbf{\Gamma}_i$ and mask vector $\mathbf{X}_i$ will be updated per training iteration.

\begin{algorithm}[t!]
\caption{Workflow of \emph{SenseCrypt}.}
\label{alg:framework}
            \parbox{7.5cm}{$\mathbf{\Gamma}_i \leftarrow$ calculate sensitivity vector as in Section \ref{sec:why:def}\;}\\
            \parbox{7.5cm}{$\alpha_i \leftarrow$ calculate encryption budget as in Section \ref{sec:encbud_bound}\;}

    \parbox{7.4cm}{$\{G_{\text{IID}}\} \leftarrow$ clustering of clients as in Section \ref{sec:sys:clustering}\;}\\
    \For{each iteration $e=1, 2, ...$}
    {
        \For{each group $G_{\emph{IID}}$ \textbf{\emph{in parallel}}}
        {
            \For{client $i \in G_{\emph{IID}}$ \textbf{\emph{in parallel}}}
            {
                $\mathbf{W}_i^e \leftarrow$  local model update as in \emph{FedAvg}\;
                \parbox{6.5cm}{$\mathbf{\Gamma}_i \leftarrow$ update sensitivity vector as in Section \ref{sec:why:def}\;}\\
                \parbox{6.5cm}{$\mathbf{X}_i \leftarrow$ calculate 
                mask vector as in Section \ref{sec:sys:cca:opt}\;}\\
                \parbox{6.5cm}{$\mathbf{W}_i^{e,*} \leftarrow$ selective HE of $\mathbf{W}_i^e$ by $\mathbf{X}_i$\;}
            }
            \parbox{7.5cm}{$\mathbf{W}^{e+1,*} \leftarrow$ aggregate $\mathbf{W}_i^{e,*}$ 
            as in \emph{FedAvg}\;}\\
            $\widehat{\mathbf{X}} \leftarrow$ union of the clients' uploaded $\mathbf{X}_i$\;
            \For{client $i \in G_{\emph{IID}}$ \textbf{\emph{in parallel}}}
            {
                $\mathbf{W}^{e+1} \leftarrow$ decryption of $\mathbf{W}^{e+1,*}$ by $\widehat{\mathbf{X}}$\;
            }
        }
    }
\end{algorithm}
\setlength{\textfloatsep}{0pt}% Remove \textfloatsep

\subsubsection{Model Aggregation without Straggling} %\label{sec:sys:agg}
Considering that \emph{FedAvg} \cite{mcmahan2017communication} has been proved as still one of the most robust FL aggregation strategies while maintaining computational simplicity \cite{jin2023fedmlhe}, especially when the data is IID, we utilize it for model aggregation per cluster without losing generality.
Moreover, since the Paillier HE supports the addition of ciphertext to plaintext \cite{paillier1999public}, unlike the traditional Selective HE methods that aggregate models by the union of clients' mask vectors, 
\emph{SenseCrypt} directly aggregates the clients' models, which are separately encrypted by their respective $\mathbf{X}_i$,
without modifying \emph{FedAvg}'s aggregation strategy.
Thus, only the decryption of the aggregation result needs the union of the clients' mask vectors (denoted as $\widehat{\mathbf{X}} = \cup_{i=1}^{N^c} \mathbf{X}_i$).

Finally, the workflow of \emph{SenseCrypt} is summarized as Algorithm \ref{alg:framework} and illustrated in Figure \ref{fig:framework}.
At Lines 1-2, each client calculates its sensitivity vector and encryption budget as in Section \ref{sec:why:def} and Section \ref{sec:encbud_bound}, respectively.
At Line 3, the server applies the clustering of clients as in Section \ref{sec:sys:clustering} to split the clients into respective clusters with IID data (denoted as $\{G_{\text{IID}}\}$).
At Lines 6-10, each client updates its sensitivity vector 
and mask vector to selectively encrypt its model parameters.
At Lines 11-12, the server aggregates the selectively encrypted models from the clients, and determines the mask vector $\widehat{\mathbf{X}}$ for decryption per cluster.
At Lines 13-14, each client uses $\widehat{\mathbf{X}}$ to decrypt the global model.
\section{Performance Evaluation} \label{sec:evaluate}

\subsection{Experimental Settings}\label{evaluate:setup}

\textbf{FL Datasets.}
We use the \texttt{CIFAR10}, \texttt{CIFAR100}, \texttt{MNIST} and \texttt{FMNIST} datasets 
for evaluations.

\textbf{Models.}
For evaluations on the \texttt{FMNIST} and \texttt{CIFAR10} datasets, we train a 
Fully-Connected Neural (FCN) network and an AlexNet \cite{krizhevsky2012imagenet}, respectively.
To evaluate effectiveness against inversion attacks, we train a ResNet-50 \cite{he2016deep} on the \texttt{CIFAR100} dataset, and a LeNet-5 \cite{lecun1998gradient} on the \texttt{MNIST} dataset.

\textbf{Implementation.}
We use PyTorch to implement \emph{SenseCrypt} on a server with 2 NVIDIA RTX 4090 GPUs, 104 Intel Xeon CPUs, and 256 GB memory.
The total number of clients is 20.
We adopt the python-paillier \cite{PythonPaillier} as the HE implementation.
The HE key size of each client is 2048.
The privacy leakage threshold $\eta_{\text{MI}}$=2.0.
For \texttt{FMNIST} and \texttt{CIFAR10}, we set \{C=0.7, B=1.3\} and \{C=0.5, B=2\}, respectively.
For details on the selection of hyperparameters B and C, please refer to Appendix section F.

\begin{figure*}[t!]
    \centering
    \begin{subfigure}[t]{0.24\linewidth}
        \centering
        \includegraphics[width = 43mm, height = 37mm]{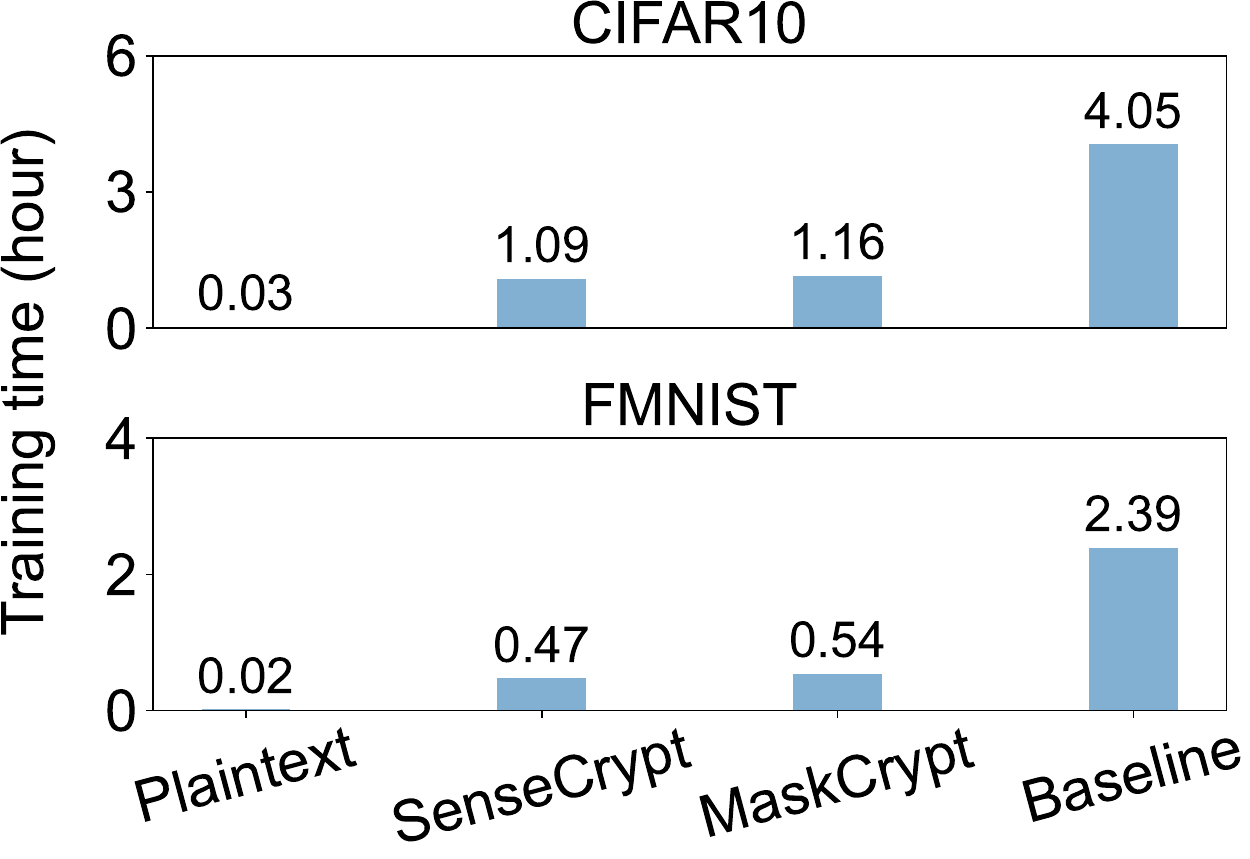}
        \caption{Training time.} \label{training_time_statistical_heterogeneity}
    \end{subfigure}
    \hspace{0.01in}
    \begin{subfigure}[t]{0.24\linewidth}
        \centering
        \includegraphics[width = 43mm, height = 37mm]{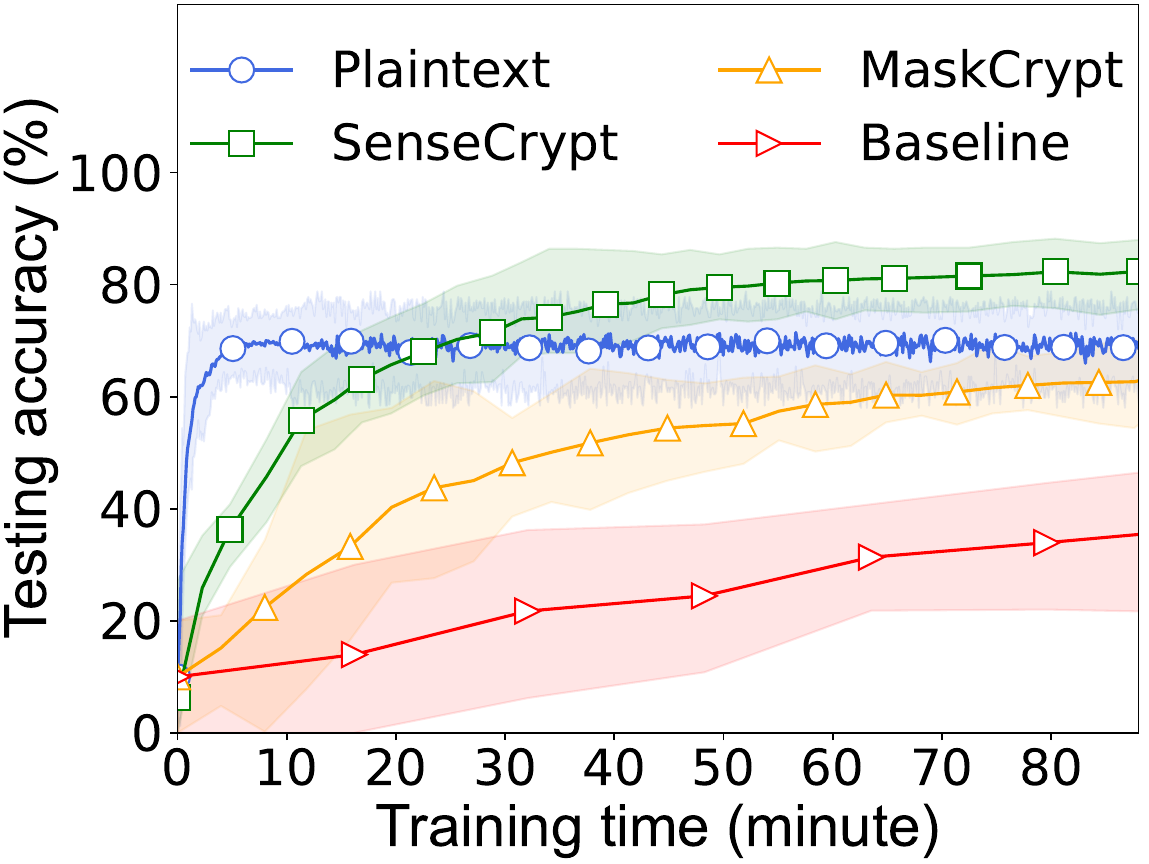}
        \caption{Performance on \texttt{CIFAR10}.} \label{testing_accuracy_statistical_heterogeneity_cifar}
    \end{subfigure}
    \hspace{0.01in}
    \begin{subfigure}[t]{0.24\linewidth}
        \centering
        \includegraphics[width = 43mm, height = 37mm]{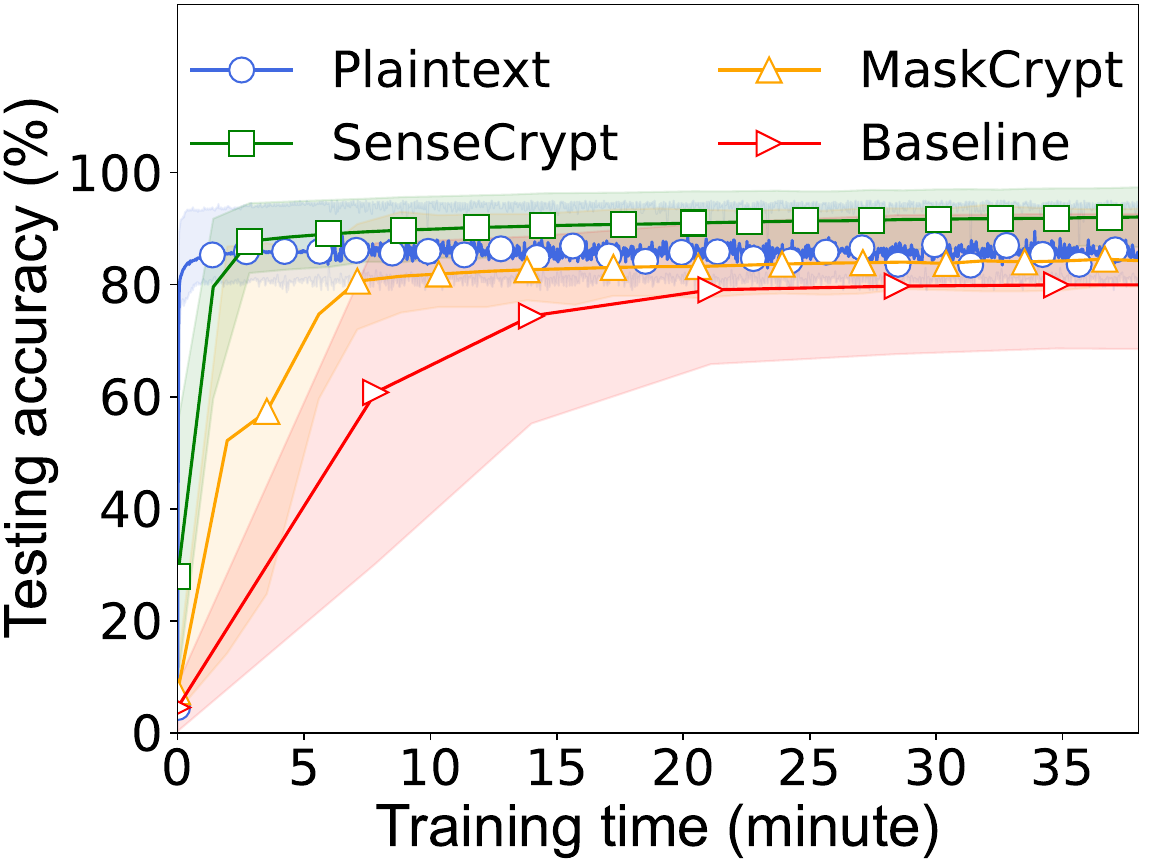}
        \caption{Performance on \texttt{FMNIST}.} \label{testing_accuracy_statistical_heterogeneity_fmnist}
    \end{subfigure}
    \hspace{0.01in}
    \begin{subfigure}[t]{0.24\linewidth}
        \centering
        \includegraphics[width = 43mm, height = 37mm]{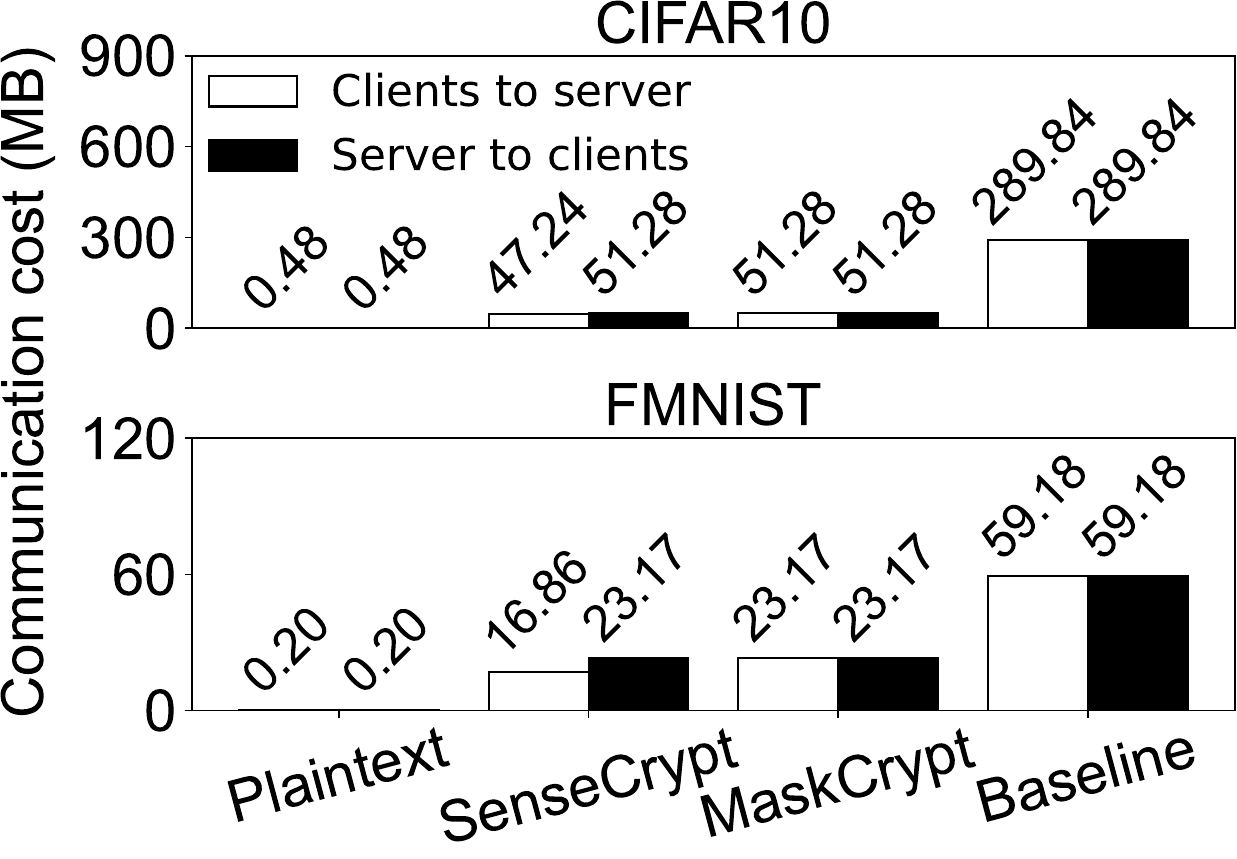}
        \caption{Communication cost.} \label{communication_cost_statistical_heterogeneity}
    \end{subfigure}
  \caption{Performance comparison in statistical heterogeneity scenario.}
  \label{newhe_stat_het}
\end{figure*}

\begin{figure*}[t!]
    \centering
    \begin{subfigure}[t]{0.24\linewidth}
        \centering
        \includegraphics[width = 43mm, height = 37mm]{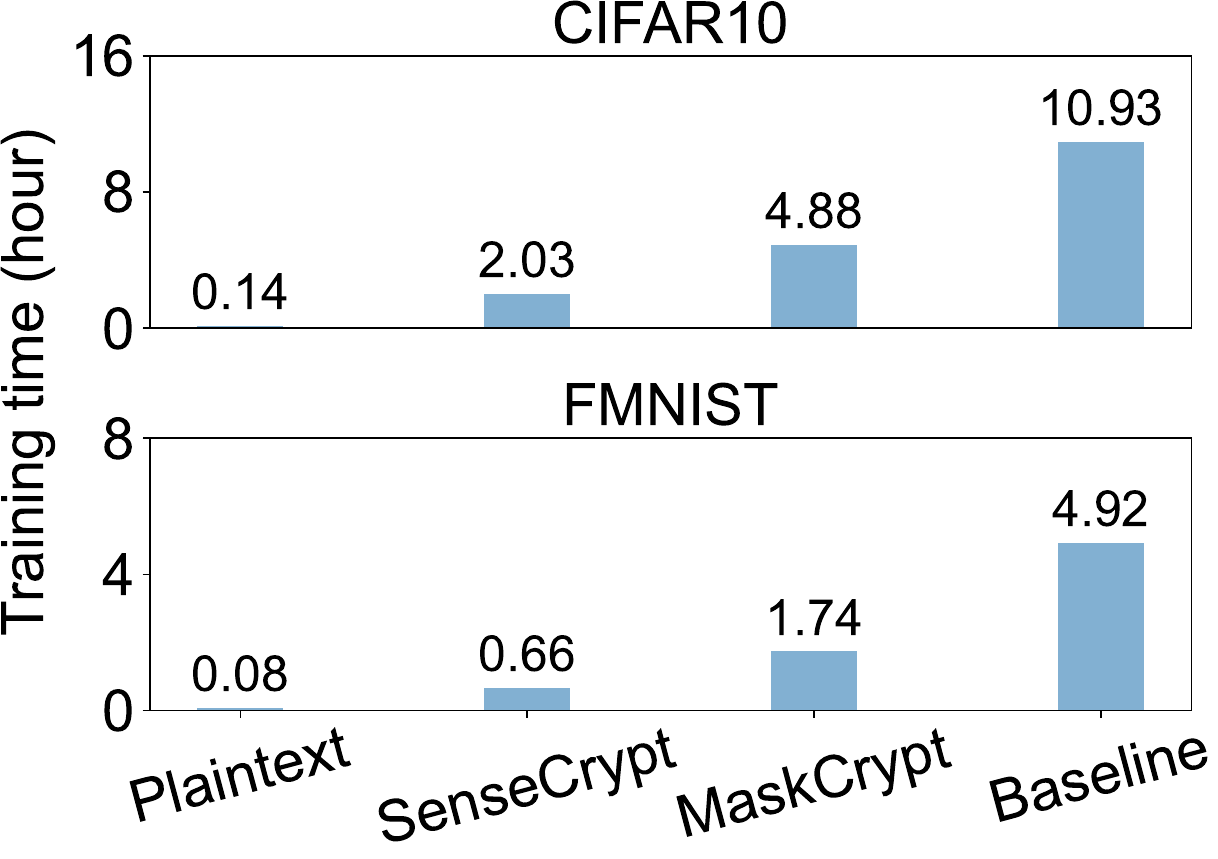}
        \caption{Training time.} \label{training_time_system_heterogeneity}
    \end{subfigure}
    \hspace{0.01in}
    \begin{subfigure}[t]{0.24\linewidth}
        \centering
        \includegraphics[width = 43mm, height = 37mm]{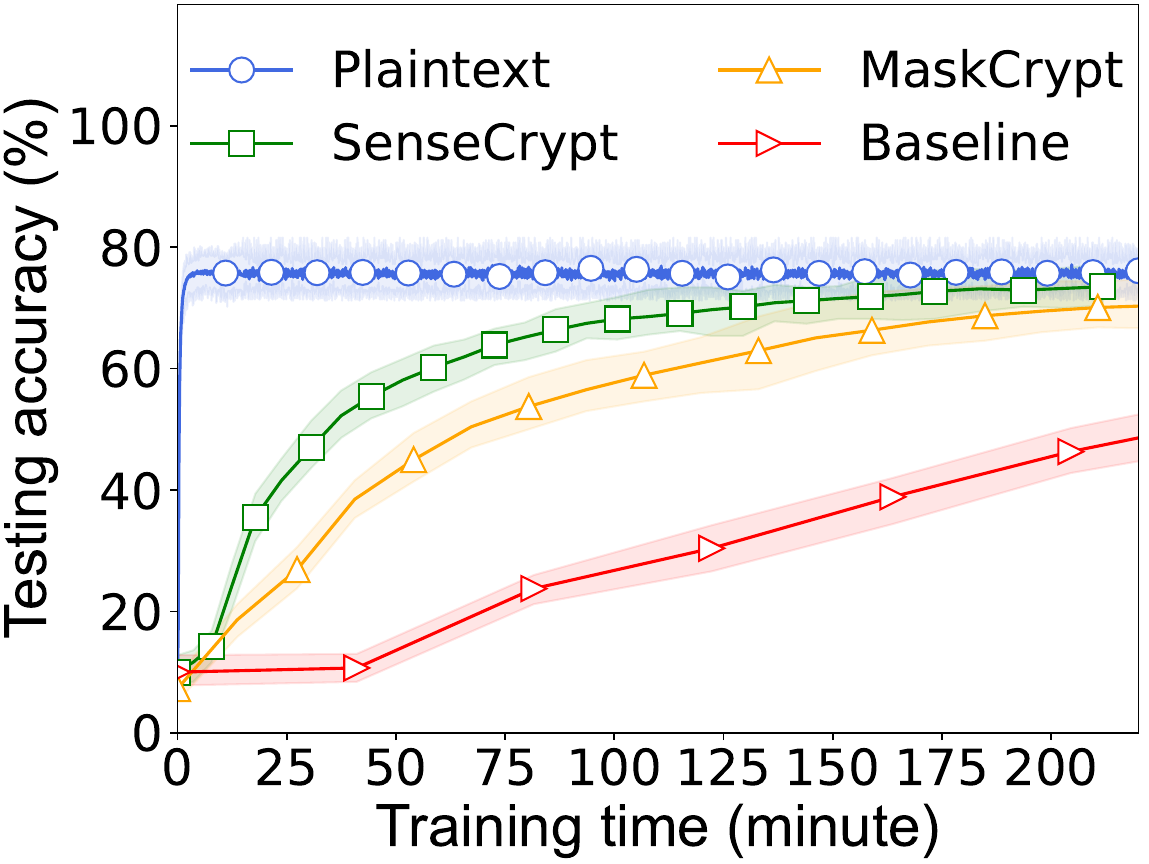}
        \caption{Performance on \texttt{CIFAR10}.} \label{testing_accuracy_system_heterogeneity_cifar}
    \end{subfigure}
    \hspace{0.01in}
    \begin{subfigure}[t]{0.24\linewidth}
        \centering
        \includegraphics[width = 43mm, height = 37mm]{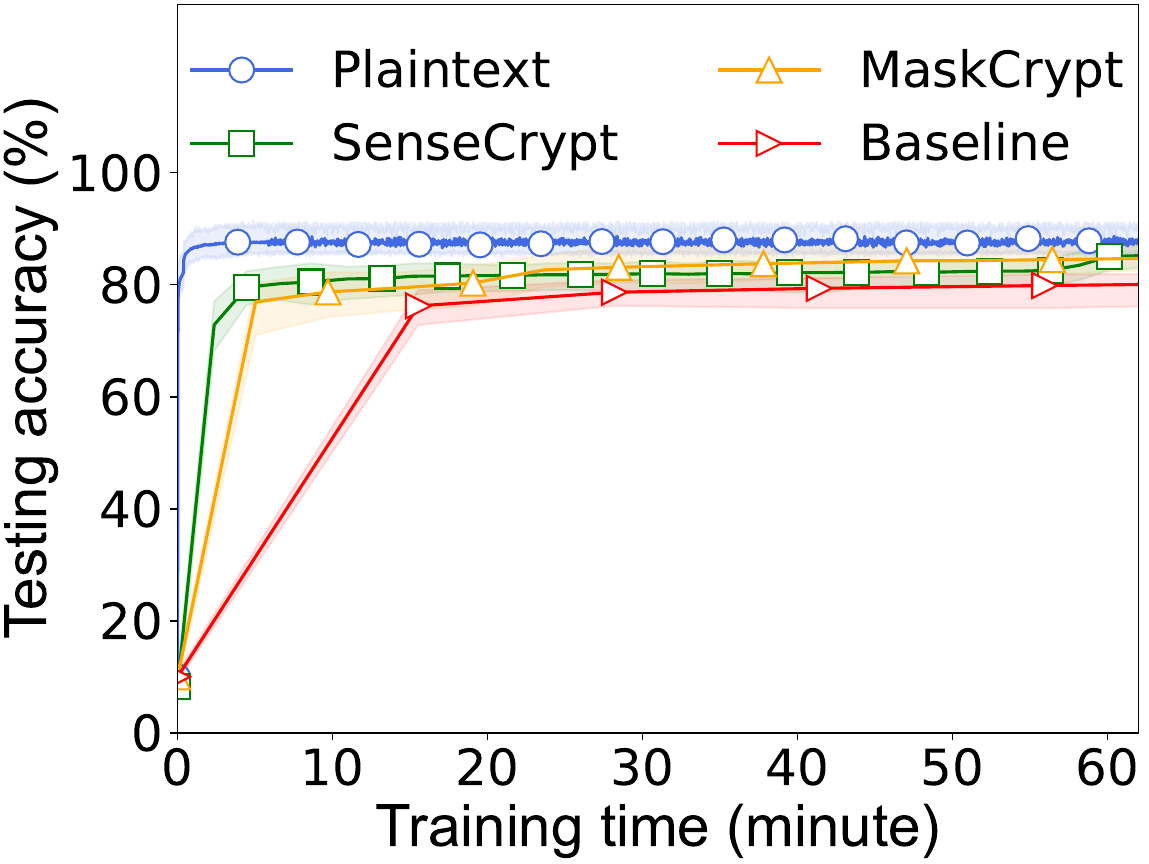}
        \caption{Performance on \texttt{FMNIST}.} \label{testing_accuracy_system_heterogeneity_fmnist}
    \end{subfigure}
    \hspace{0.01in}
    \begin{subfigure}[t]{0.24\linewidth}
        \centering
        \includegraphics[width = 43mm, height = 37mm]{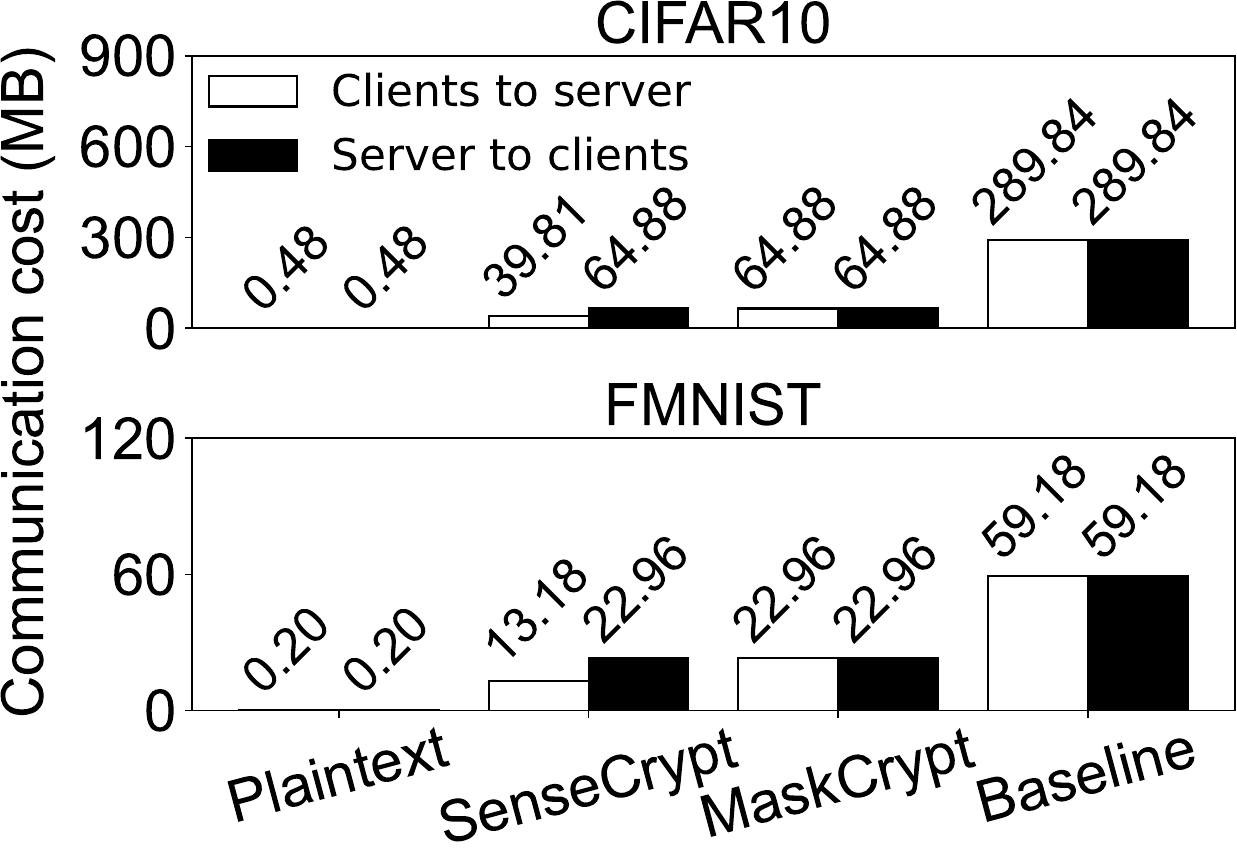}
        \caption{Communication cost.} \label{communication_cost_system_heterogeneity}
    \end{subfigure}
  \caption{Performance comparison in system heterogeneity scenario.}
  \label{newhe_sys_het}
\end{figure*}

\begin{figure*}[t!]
    \centering
    \begin{subfigure}[t]{0.24\linewidth}
        \centering
        \includegraphics[width = 43mm, height = 37mm]{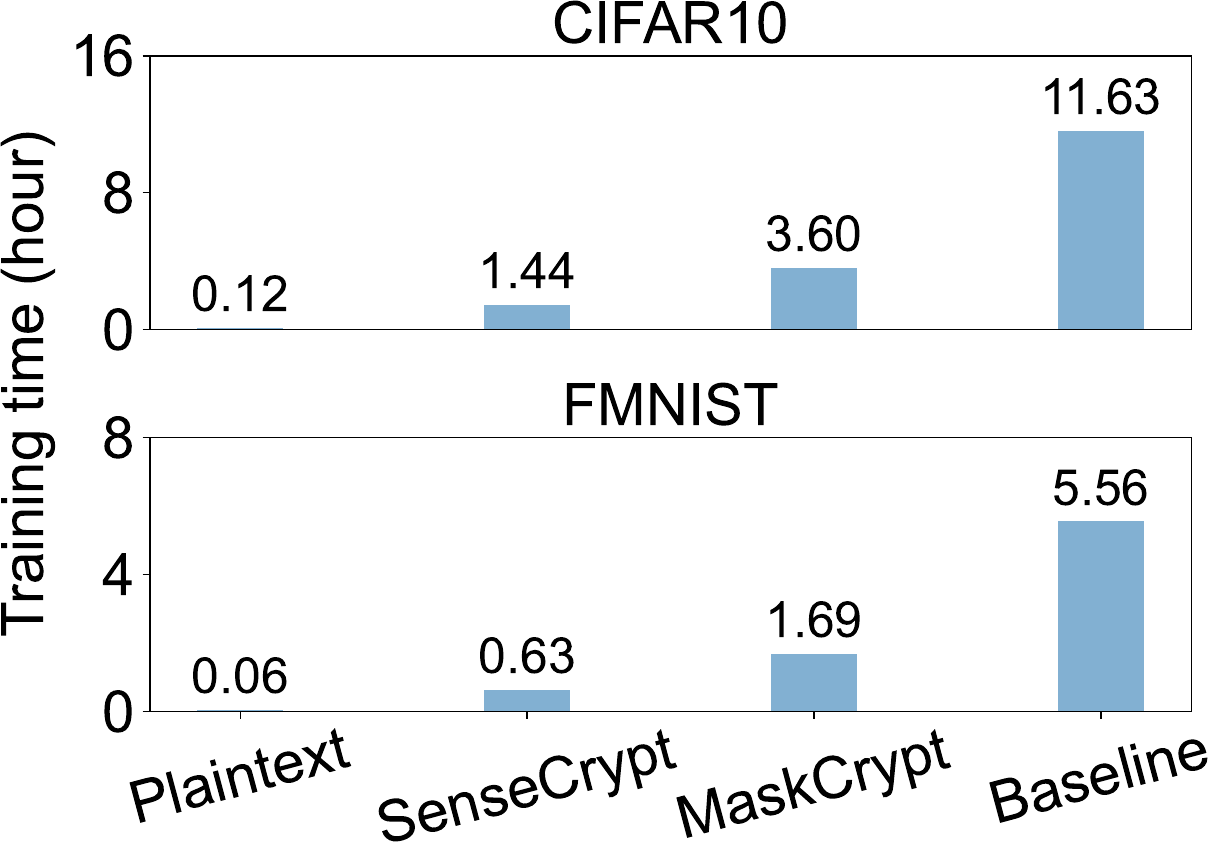}
        \caption{Training time.} \label{training_time_system_statistical_heterogeneity}
    \end{subfigure}
    \hspace{0.01in}
    \begin{subfigure}[t]{0.24\linewidth}
        \centering
        \includegraphics[width = 43mm, height = 37mm]{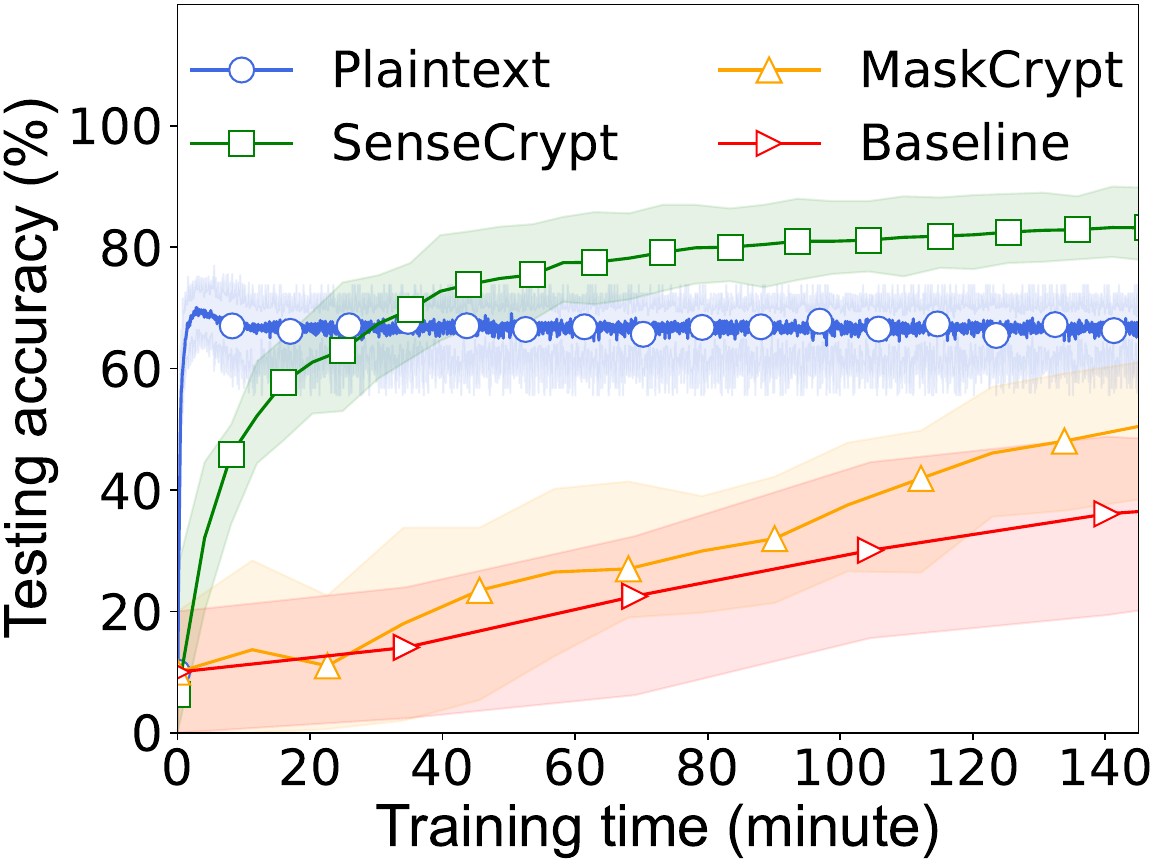}
        \caption{Performance on \texttt{CIFAR10}.} \label{testing_accuracy_system&statistical_heterogeneity_cifar}
    \end{subfigure}
    \hspace{0.01in}
    \begin{subfigure}[t]{0.24\linewidth}
        \centering
        \includegraphics[width = 43mm, height = 37mm]{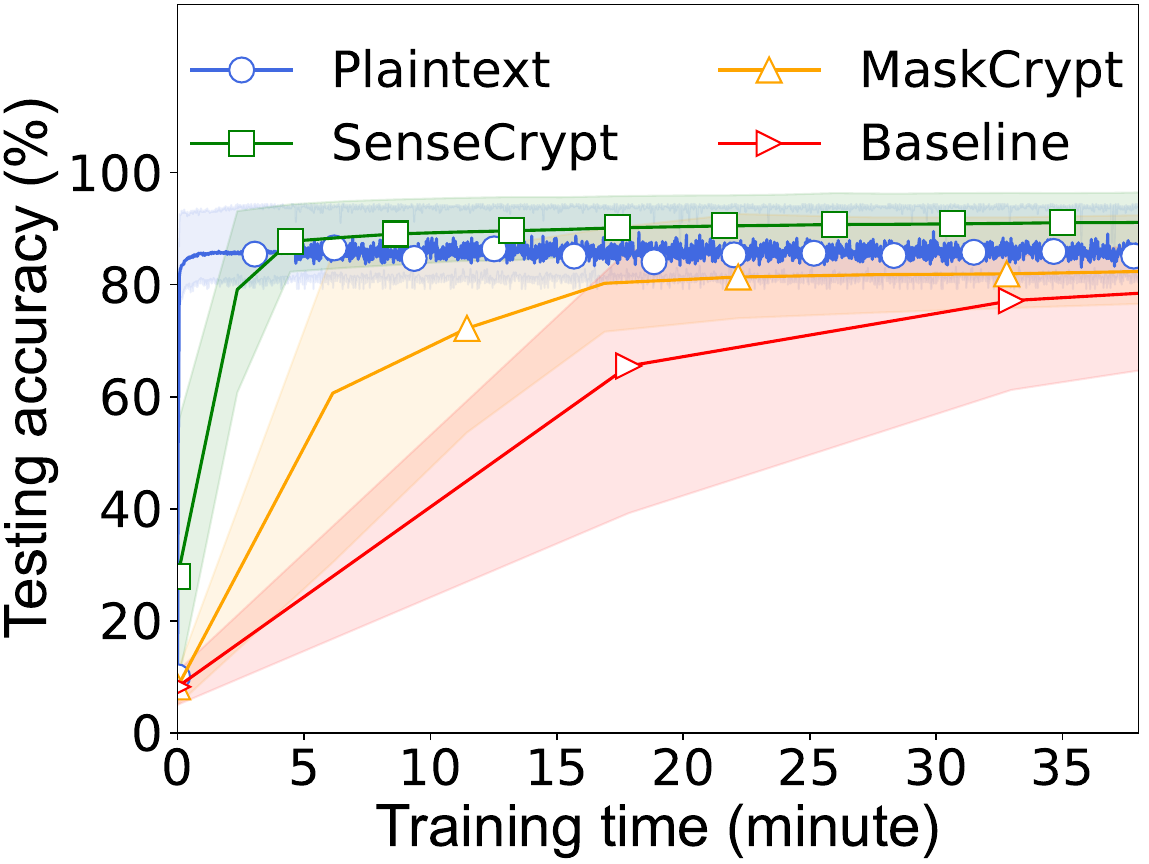}
        \caption{Performance on \texttt{FMNIST}.} \label{testing_accuracy_system&statistical_heterogeneity_fmnist}
    \end{subfigure}
    \hspace{0.01in}
    \begin{subfigure}[t]{0.24\linewidth}
        \centering
        \includegraphics[width = 43mm, height = 37mm]{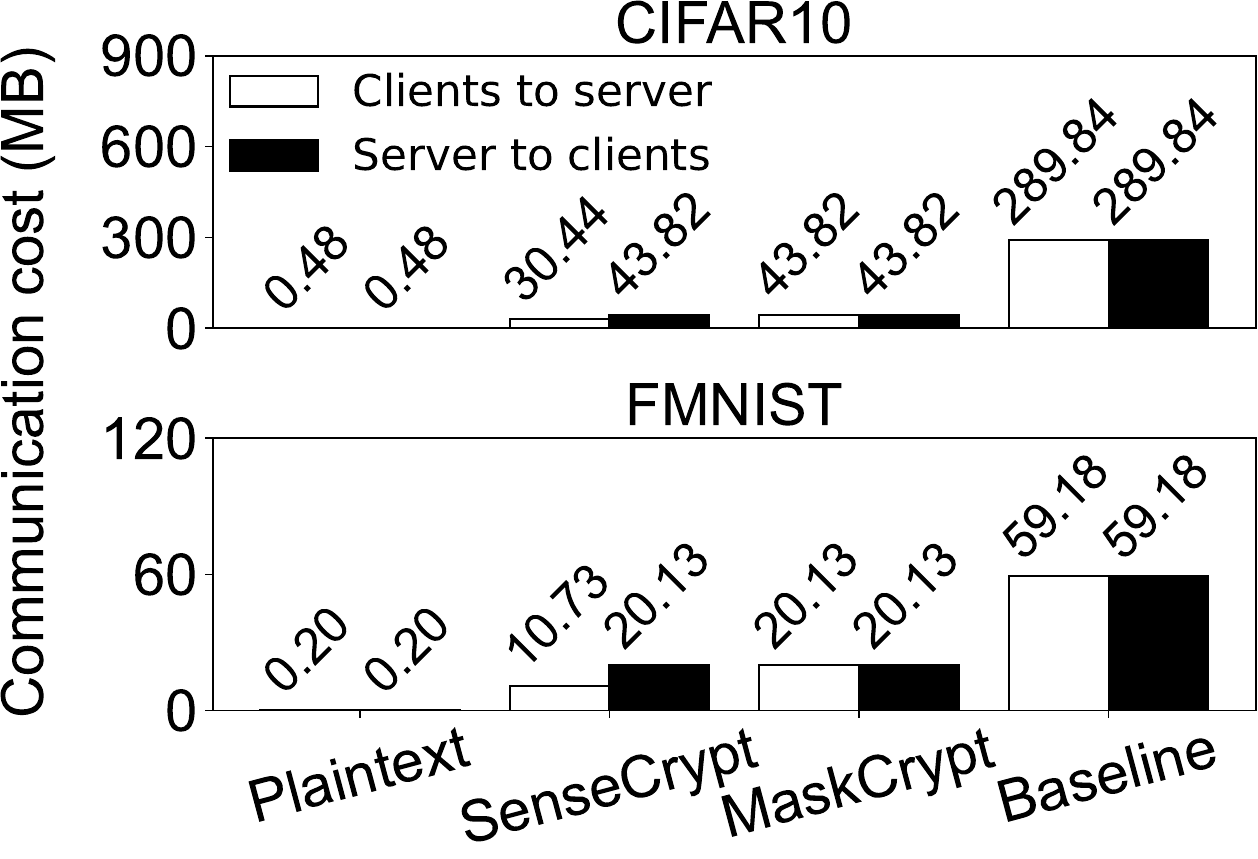}
        \caption{Communication cost.} \label{communication_cost_system_statistic_heterogeneity}
    \end{subfigure}
  \caption{Performance comparison in statistical \& system heterogeneity scenario.}
  \label{newhe_stat_sys_het}
\end{figure*}

\textbf{Experiment Scenarios.}
We design three scenarios:
\begin{itemize}[
fullwidth,
topsep=0pt,
      itemindent=1em,
]

    \item \textbf{Statistical Heterogeneity Scenario.}
    We evenly split the clients into 4 categories 
    and ensure the clients' data is IID within the same category, but Non-IID across different categories.
    To exclusively evaluate the impact of statistical heterogeneity, each client has 5\% of overall data, a bandwidth of 50 MBps and 32 CPUs.

    \item \textbf{System Heterogeneity Scenario.}
    We first evenly split the clients into 4 categories.
    Then, 
    1) for bandwidth heterogeneity, we assign 50, 45, 40, 35 and 30 MBps to the 5 clients per category, respectively,
    2) for computation speed heterogeneity, we assign 24, 16, 12, 10 and 8 CPUs to the 5 clients per category, respectively.
    To exclusive evaluation of system heterogeneity, all the clients hold IID training data.

    \item \textbf{Statistical \& System Heterogeneity Scenario.}
    We divide the clients into 4 categories, each containing 5 clients, and adopt the same settings as in the above scenarios.
\end{itemize}

\textbf{Comparison Methods.}
We compare 
\emph{SenseCrypt} with a representative Selective HE method (denoted as \emph{MaskCrypt}) \cite{hu2024maskcrypt}, 
\emph{FedAvg} with ciphertext encrypted by the Paillier HE method (denoted as \emph{Baseline}) \cite{PythonPaillier}, and \emph{FedAvg} with plaintext (denoted as \emph{Plaintext}) \cite{mcmahan2017communication}.
Specifically, \emph{MaskCrypt} utilizes the change of gradient value to select the model parameters that need encryption.
\emph{Baseline} utilizes the stock implementation of python-paillier \cite{PythonPaillier} to encrypt each model parameter.

\subsection{Experimental Results}\label{overall results}

\subsubsection{Performance in Heterogeneity Scenarios} \label{eval:stathet}

Figures \ref{newhe_stat_het}, \ref{newhe_sys_het} and \ref{newhe_stat_sys_het} show the evaluation results
of different methods 
in the three scenarios.
For fairness, the union of masks 
in \emph{SenseCrypt} has the same size as that 
in \emph{MaskCrypt}.

Figures \ref{training_time_statistical_heterogeneity}, \ref{training_time_system_heterogeneity} and \ref{training_time_system_statistical_heterogeneity} show the training time of the methods over 30 iterations in the three scenarios.
We can see that \emph{Plaintext} consistently results in the shortest training time as it spends no time on HE, whereas \emph{Baseline} consistently results in the longest training time as it encrypts each model parameter. %, which significantly increases the total HE time cost.
In comparison, \emph{MaskCrypt} greatly shortens the training time.
This is because that \emph{MaskCrypt} only selectively encrypts partial model parameters by sensitivity. %instead of encrypting every single one of them.
However, \emph{MaskCrypt} still consumes \textgreater33$\times$ and \textgreater20$\times$ more time than that of \emph{Plaintext} on \texttt{CIFAR10} and \texttt{FMNIST}, respectively, in both system heterogeneity and statistical \& system heterogeneity scenarios.
This is because that \emph{MaskCrypt} cannot assign encryption masks in accordance with the clients' system capabilities, thereby deteriorates the straggler problem in these scenarios.
In contrast, \emph{SenseCrypt} reduces the training time by 58.4\%$\sim$62.7\% and 81.4\%$\sim$88.7\% as compared to \emph{MaskCrypt} and \emph{Baseline}, respectively, in Figures \ref{training_time_system_heterogeneity} and \ref{training_time_system_statistical_heterogeneity}.
Even in the statistical heterogeneity scenario where clients share the same system capabilities (Figure \ref{training_time_statistical_heterogeneity}), \emph{SenseCrypt} requires shorter training time than \emph{MaskCrypt}.
This is primarily due to the adaptive selection of model parameters for encryption in \emph{SenseCrypt}, which not only considers the clients' system capabilities, but also adapts to their Non-IID data via client clustering. %based on model parameter sensitivity.
We also notice that the training time of \emph{MaskCrypt} and \emph{SenseCrypt} in Figure \ref{training_time_system_statistical_heterogeneity} is slightly shorter than that in Figure \ref{training_time_system_heterogeneity}.
The primary reason is that, in \emph{SenseCrypt}, the number of clients per cluster (i.e., 5 clients sharing IID data) in the statistical \& system heterogeneity scenario is smaller than that in the system heterogeneity scenario (i.e., 20 clients sharing IID data), which causes the union of encryption masks per cluster shrinks in size.
For fairness, the mask size in \emph{MaskCrypt} also shrinks correspondingly.

Figures \ref{testing_accuracy_statistical_heterogeneity_cifar}, \ref{testing_accuracy_statistical_heterogeneity_fmnist}, \ref{testing_accuracy_system_heterogeneity_cifar}, \ref{testing_accuracy_system_heterogeneity_fmnist}, \ref{testing_accuracy_system&statistical_heterogeneity_cifar} and \ref{testing_accuracy_system&statistical_heterogeneity_fmnist} show the testing accuracy of each client's trained model of different methods in the three scenarios.
Each curve represents the mean testing accuracy of all the clients, and the shaded areas surrounding the curve represent the fluctuation range of testing accuracies among the clients.
We can see that despite the fast convergence rate of \emph{Plaintext}, 
the clients' testing accuracy after convergence is significantly lower than that of \emph{SenseCrypt} in the statistical heterogeneity scenario and the statistical \& system heterogeneity scenario.
Similarly, such model performance degradation in scenarios with high statistical heterogeneity can also be observed in \emph{MaskCrypt} and \emph{Baseline}.
The reason is that these methods cannot adapt to the clients' Non-IID data, causing the global models trained via FL to fail to converge to satisfactory performance levels.
In contrast, \emph{SenseCrypt} can always achieve normal model performance levels as on IID data, while enjoying approximate convergence rate as \emph{Plaintext}, especially on \texttt{FMNIST}.
This is majorly due to the client clustering, which ensures that the clients per group have IID data, and the adaptive selection of model parameters  for encryption, which prevents convergence delay caused by straggling.

\begin{figure*}[t!]
    \centering
    \begin{minipage}[t]{0.24\linewidth}
    \centering
  \includegraphics[width = 43mm, height = 37mm]{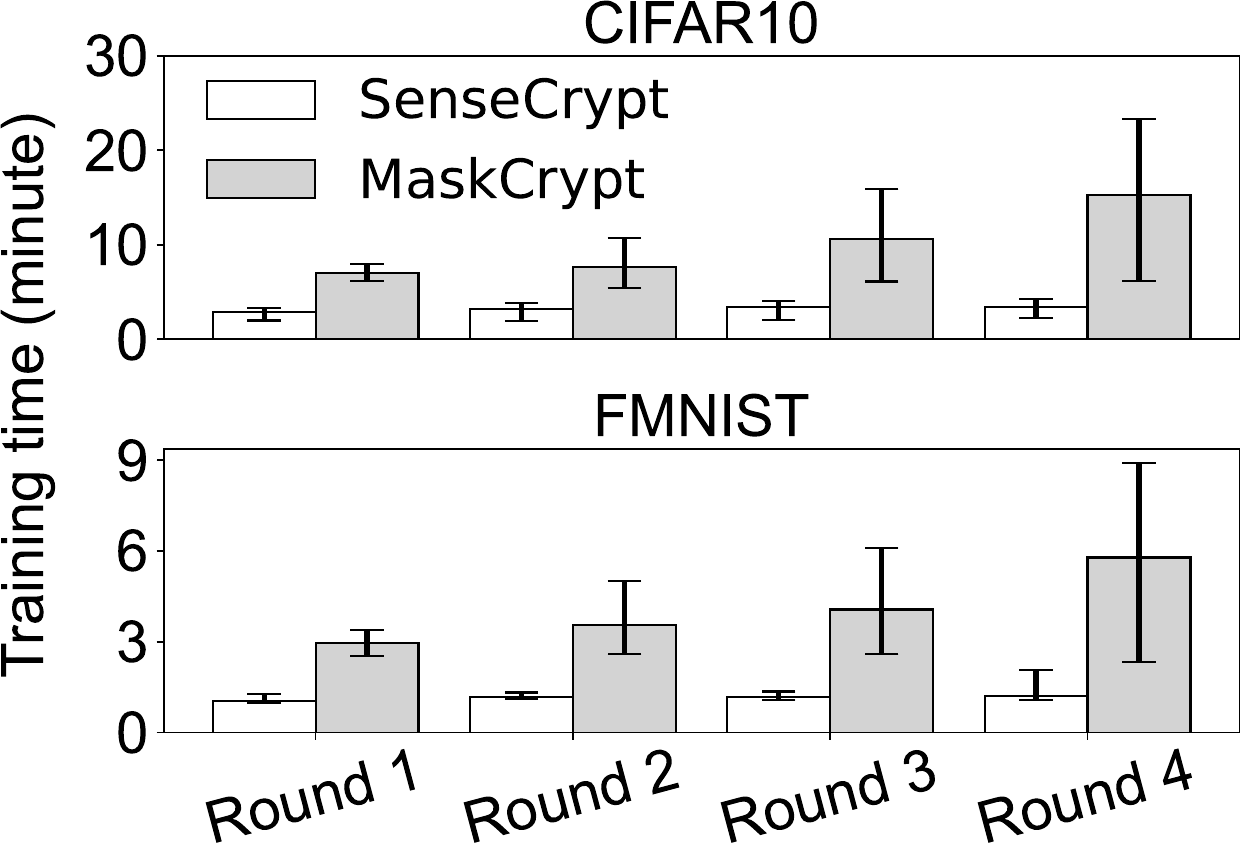}
  \caption{Training time under system heterogeneity.}
   \label{eval_train_time}
\end{minipage}
\hspace{0.01in}
\begin{minipage}[t]{0.24\linewidth}
    \centering
  \includegraphics[width = 43mm, height = 37mm]{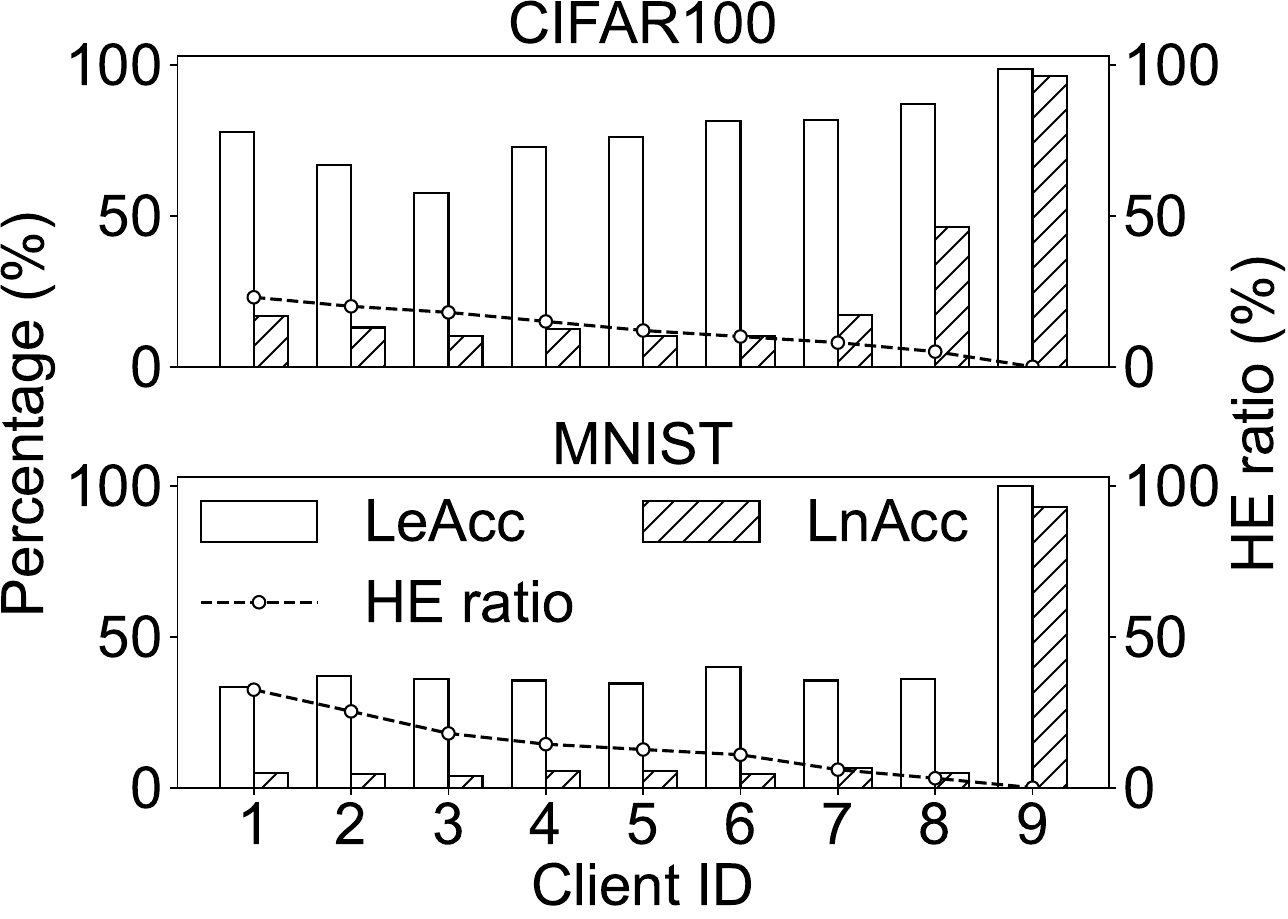}
  \caption{Attack results under extreme system heterogeneity.}
   \label{eval_attack_rate_sys_het}
   \end{minipage}
\hspace{0.01in}
    \begin{minipage}[t]{0.24\linewidth}
    \centering
  \includegraphics[width = 43mm, height = 37mm]{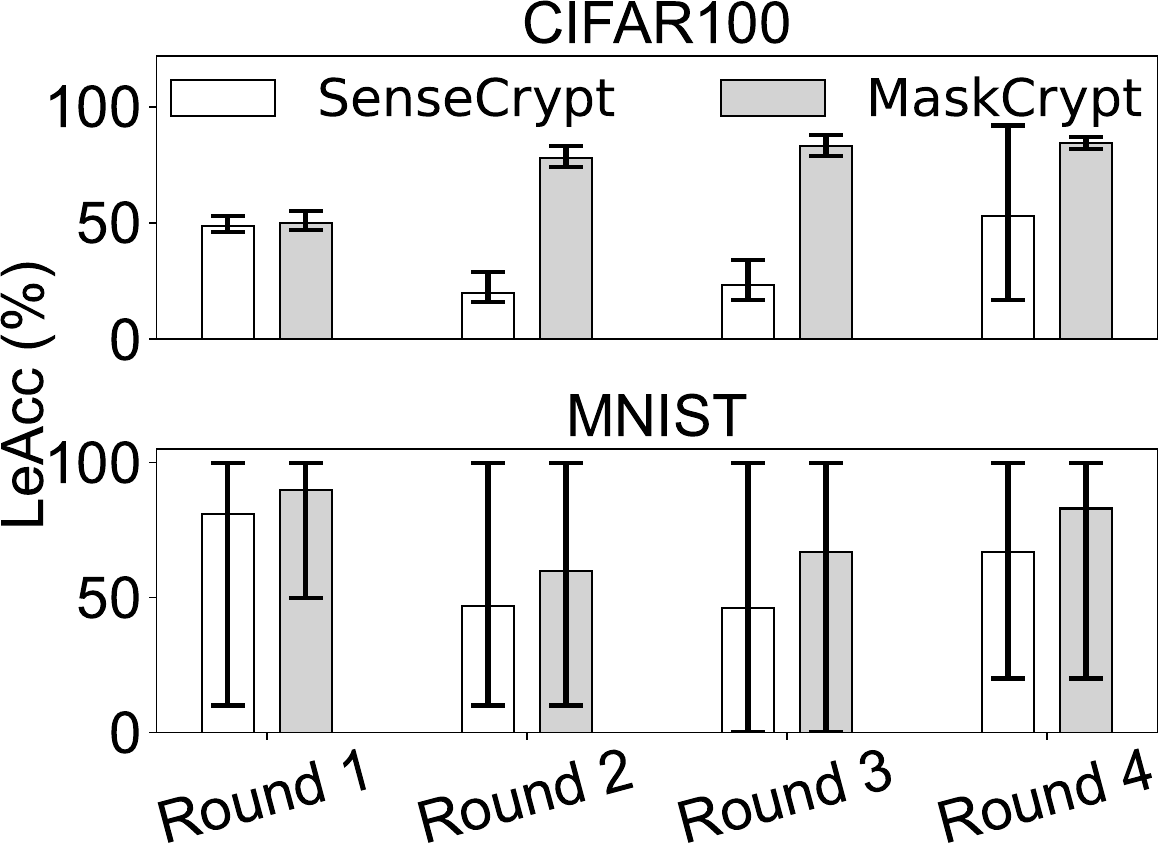}
  \caption{LeAcc under statistical heterogeneity.}
   \label{eval_attack_rate_stat_het_leacc}
\end{minipage}
 \hspace{0.01in}
\begin{minipage}[t]{0.24\linewidth}
    \centering
  \includegraphics[width = 43mm, height = 37mm]{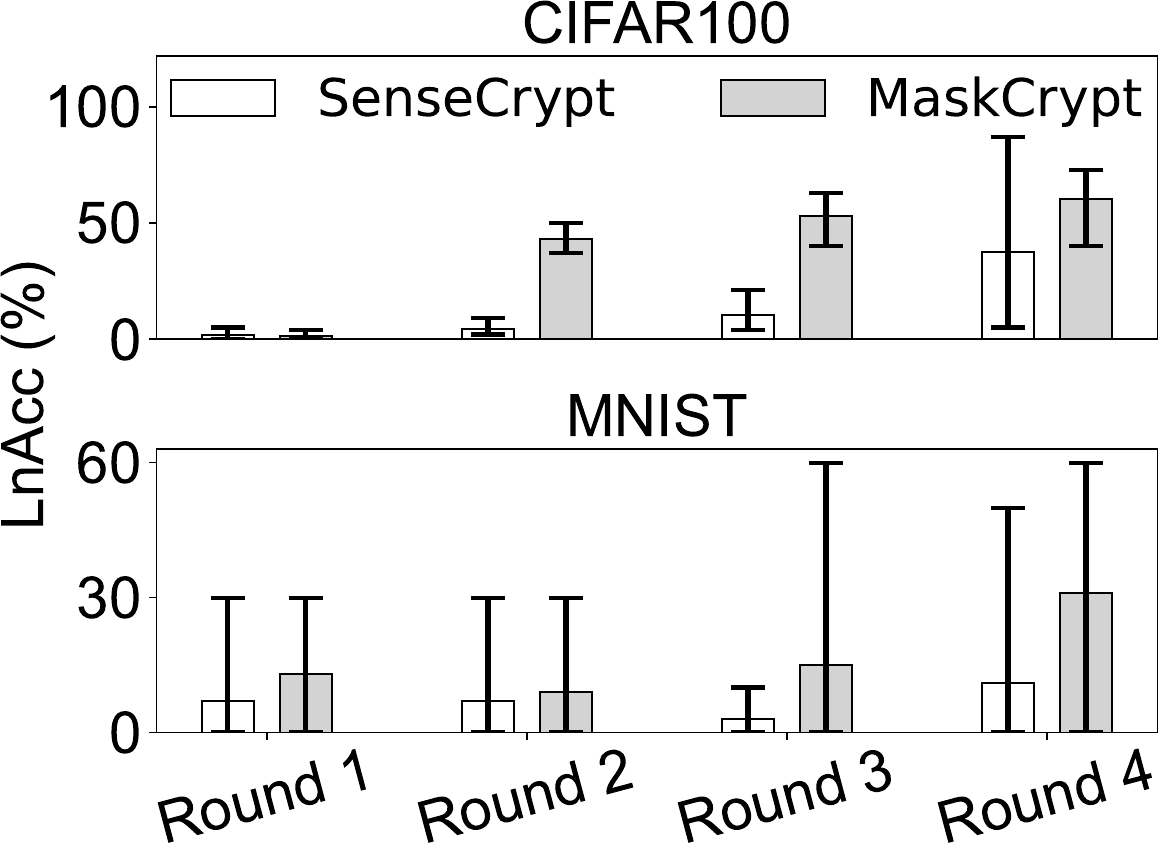}
  \caption{LnAcc under statistical heterogeneity.}
   \label{eval_attack_rate_stat_het_lnacc}
   \end{minipage}
\end{figure*}

Figures \ref{communication_cost_statistical_heterogeneity}, \ref{communication_cost_system_heterogeneity} and \ref{communication_cost_system_statistic_heterogeneity} show the communication cost of different methods incurred in one iteration in the three scenarios.
The upstream communication cost (i.e., clients $\rightarrow$ server) and the downstream communication cost (i.e., server $\rightarrow$ clients) are separately illustrated.
Given that the ciphertext generated by the Paillier HE method increases in size by more than $64\times$ as compared to the plaintext \cite{zhang2020batchcrypt}, the fewer model parameters are encrypted in a method, the lower communication cost it will render.
We can see that the results are generally consistent with the training time due to the same aforementioned reasons.
Note that unlike the comparison methods, the upstream communication cost of \emph{SenseCrypt} is always much lower than its downstream communication cost.
This is because that \emph{SenseCrypt} uses the adaptively determined mask to encrypt partial model parameters of respective clients, thereby further reducing the communication cost.

\subsubsection{Cost Efficiency under Varying Heterogeneity Degrees} \label{eval:cost_efficiency}

We further evaluate cost efficiency in terms of training time under varying system heterogeneity degrees.
Specifically, we conduct FL training on 5 clients for 4 rounds and gradually increase the variance in the number of CPUs as: Round 1 ([32, 30, 28, 26, 24]), Round 2 ([32, 28, 24, 20, 16]), Round 3 ([32, 24, 18, 16, 12]) and Round 4 ([32, 16, 12, 10, 8]).
To avoid the impact of statistical heterogeneity, we set that the clients have IID training data, 20\% of overall data samples, and a bandwidth of 50 MBps. 
For fairness, the size of the union of masks is the same in \emph{SenseCrypt} and \emph{MaskCrypt}.
Then, we measure each client's training time per iteration.

Figure \ref{eval_train_time} shows the minimum, mean and maximum values of all the measured results under each system heterogeneity degree on \texttt{CIFAR10} and \texttt{FMNIST}.
We can see that as the system heterogeneity degree increases, the mean and fluctuation of the clients' training time in \emph{MaskCrypt} also significantly rise on both datasets.
In contrast, the training time of \emph{SenseCrypt} remains nearly static across different system heterogeneity degrees, primarily due to the adaptive partial encryption of model parameters in accordance with the clients' system capabilities.
We hence conclude that \emph{SenseCrypt} can tolerate varying system heterogeneity degrees.

\subsubsection{Effectiveness against Inversion Attacks} \label{eval:effect_safe}

Intuitively, under extremely high system heterogeneity, the slowest client's encryption mask determined by \emph{SenseCrypt} might be very small, which may affect its security.
To evaluate the effectiveness of \emph{SenseCrypt} against inversion attacks in such cases, we further increase the system heterogeneity degree and launch inversion attacks to reconstruct each client's private data.
Specifically, we conduct another round of FL training on 9 clients. 
The numbers of CPUs of Clients 1-8 follow [32, 24, 16, 12, 10, 8, 2, 1].
For reference, Client 9 trains without HE.
The other settings are the same as in Section \ref{eval:cost_efficiency}.
Then, we launch the instance-wise Labels Restoration from Gradients (iLRG) attack \cite{ma2023instance} on each client, which utilizes the batch-averaged gradients 
to reconstruct each data sample and its 
label per batch. %via solely using the batch-averaged gradients in FL training.
To evaluate the general effectiveness, 
the 1\textsuperscript{st} attack is launched on ResNet-50 with a batch size of 200, a model depth (i.e., number of model layers) of 50, and the \texttt{CIFAR100} dataset.
The 2\textsuperscript{nd} attack is launched on LeNet-5 with a batch size of 30, a model depth of 7, and the \texttt{MNIST} dataset.
Finally, we calculate the Label existence Accuracy (LeAcc), which measures the accuracy score for predicting label existences, and the Label number Accuracy (LnAcc), which measures the accuracy score for predicting the number of instances per class, of the reconstructed results as in \cite{ma2023instance}.

Figure \ref{eval_attack_rate_sys_het} shows the measured results
of each client.
We find that among Clients 1-8, the HE ratio decreases along with the decrement in the number of CPUs to avoid straggling.
Interestingly, no matter how low the HE ratio is (as low as 5.7\% on \texttt{CIFAR100} and 3.1\% on \texttt{MNIST}), LeAcc remains almost static and LnAcc is significantly lower than that of Client 9 (no encryption). %on both datasets.
This confirms that the redundancy of model parameters allows the adaptive selective HE of model parameters while ensuring each client's security under extremely high system heterogeneity.

We additionally launch iLRG attacks under varying statistical heterogeneity degrees.
Specifically, we conduct FL training on 10 clients for 4 rounds: 
1) in the 1\textsuperscript{st} round, each client has equal splits of all data classes (i.e., strictly IID); 
2) in the 2\textsuperscript{nd}, 3\textsuperscript{rd} and 4\textsuperscript{th} rounds, each client has all the data samples of 5, 2 and 1 classes, respectively.
For the rest classes, each client has one data sample per class.
For fairness, both \emph{SenseCrypt} and \emph{MaskCrypt} encrypt 5.0\% and 3.1\% of the model parameters on \texttt{CIFAR100} and \texttt{MNIST}, respectively. 

Figures \ref{eval_attack_rate_stat_het_leacc} and \ref{eval_attack_rate_stat_het_lnacc} show the minimum, mean and maximum values of the measured results of different methods
per round.
Note that the exceptionally high LeAcc in Round 1 of Figure \ref{eval_attack_rate_stat_het_leacc} is caused by the label prediction logic of iLRG, which infers that all labels exist by default \cite{ma2023instance}, thereby favoring the case with strictly IID data.
We can see that along with the increase of Non-IID degree, the LeAcc and LnAcc of \emph{MaskCrypt} significantly rise, while those of \emph{SenseCrypt} almost remain static.
This is primarily because that under a more Non-IID degree, the union of sensitive (important) model parameters that need encryption becomes larger, which cannot be fully covered by the shared encryption mask of \emph{MaskCrypt}.
While in \emph{SenseCrypt}, the clustering of clients with IID data avoids the expansion of important model parameters that need encryption, thereby ensuring sufficient encryption of the important model parameters per cluster under varying Non-IID degrees.

\section{Conclusion}~\label{sec:conclusion}
We propose \emph{SenseCrypt}, a Selective HE framework that clusters clients with IID data by model parameter sensitivity, assigns straggler-free encryption budget, and adaptively balances model security and HE overhead for cross-device FL clients with heterogeneous data and system capabilities.
Our experiments conducted in multiple heterogeneity scenarios showed that compared with the state-of-the-art, \emph{SenseCrypt} achieves normal model accuracy as on IID data, while reducing training time by 58.4\%$\sim$88.7\%, 
and ensuring model security against the state-of-the-art inversion attacks.

\bibliography{ref}

\clearpage

\appendix

\section{Notations}
\label{app:notations}

\begin{table}[h]
\centering
\small
\begin{tabular}{>{\centering\arraybackslash}m{1.6cm}|>{\arraybackslash}m{6.0cm}}
\hline
\textbf{Notation} & \textbf{Description} \\ 
\hline
$\mathbf{W}$         & Model parameters of a neural network \\
\hline
$\mathbf{W}_{-\mathbf{w}}$     & Model parameters with $\mathbf{w}$ zeroed-out \\
\hline
$\mathcal{L}(\mathbf{W})$ & Model loss function \\
\hline
$\nabla_{\mathbf{W}}\mathcal{L}(\mathbf{W})$   & Gradients of the loss function with respect to $\mathbf{W}$ \\
\hline
$\mathbf{\Gamma}(\mathbf{w})$       & Sensitivity of a subset of model parameters $\mathbf{w}$ \\
\hline
$\mathbf{\Gamma}_i$       & Sensitivity vector of the $i$-th client \\
\hline
$I(\mathbf{W}; \mathbf{W}_{-\mathbf{w}})$   & The mutual information between $\mathbf{W}$ and $\mathbf{W}_{-\mathbf{w}}$  \\
\hline
$\gamma_k^i$       & The $k$-th element in $\mathbf{\Gamma}_i$ \\
\hline
$\alpha_i$        & Encryption budget of the $i$-th client \\
\hline
$r_i$             & Bandwidth of the $i$-th client \\
\hline
$v_i$             & CPU clock speed of the $i$-th client \\
\hline
$N^c$             & Number of clients in a cluster \\
\hline
$N^{\mathbf{w}}$  & Total number of model parameters \\
\hline
$N^p$             & Number of input pixels in a local batch of training data \\
\hline
$N^b$             & Batch size \\
\hline
$n$               & Number of pixels in a data sample \\
\hline
$\beta_l$         & Weight of objective function $l$ in the optimization problem  \\
\hline
$\mathbf{X}_i$ & Encryption mask vector of the $i$-th client \\
\hline
$\widehat{\mathbf{X}}$ & Union of clients' encryption mask vectors in a cluster \\
\hline
$G_{\text{IID}}$  & Set of clusters with IID data \\ 
\hline
\end{tabular}
\caption{Table of main notations.}
\label{tab:notations}
\end{table}
\section{Paillier Homomorphic Encryption}
\label{app:paillier}

The Paillier cryptosystem \cite{paillier1999public} is a probabilistic asymmetric encryption scheme classified as a Partial Homomorphic Encryption (HE) system. 
Its security relies on the decisional composite residuosity assumption (DCRA), which posits the computational infeasibility of distinguishing random elements modulo $n^2$ from the $n$-th residues, where $n = pq$ is an RSA modulus with large primes $p$ and $q$. 
Paillier satisfies the additive homomorphic property: 
$$\text{Enc}(m_1) \cdot \text{Enc}(m_2) \equiv \text{Enc}(m_1 + m_2 \mod n) \mod n^2,$$ 
which makes it particularly suitable for Federated Learning scenarios where encrypted local model updates need to be aggregated without decryption.

\section{Straggler Problem of Existing Selective HE Methods in System Heterogeneity Scenarios} \label{ana:sys_het}
\label{app:system_heterogeneity}

In existing Selective HE methods \cite{jin2023fedmlhe,hu2024maskcrypt}, all clients use 
the same encryption mask to encrypt 
model parameters.
However, in scenarios with high system capability heterogeneity, 
the clients with the lowest capability may need substantially more time to encrypt the masked model parameters per training iteration, and always delay the completion of FL training (i.e., the \emph{straggler problem} \cite{chai2020tifl}).
To confirm this conjecture, we vary the encryption budget of the clients, 
and measure 
their HE time costs (encryption plus decryption time).

\begin{figure}[t!]
    \centering
    \includegraphics[width=0.75\linewidth]{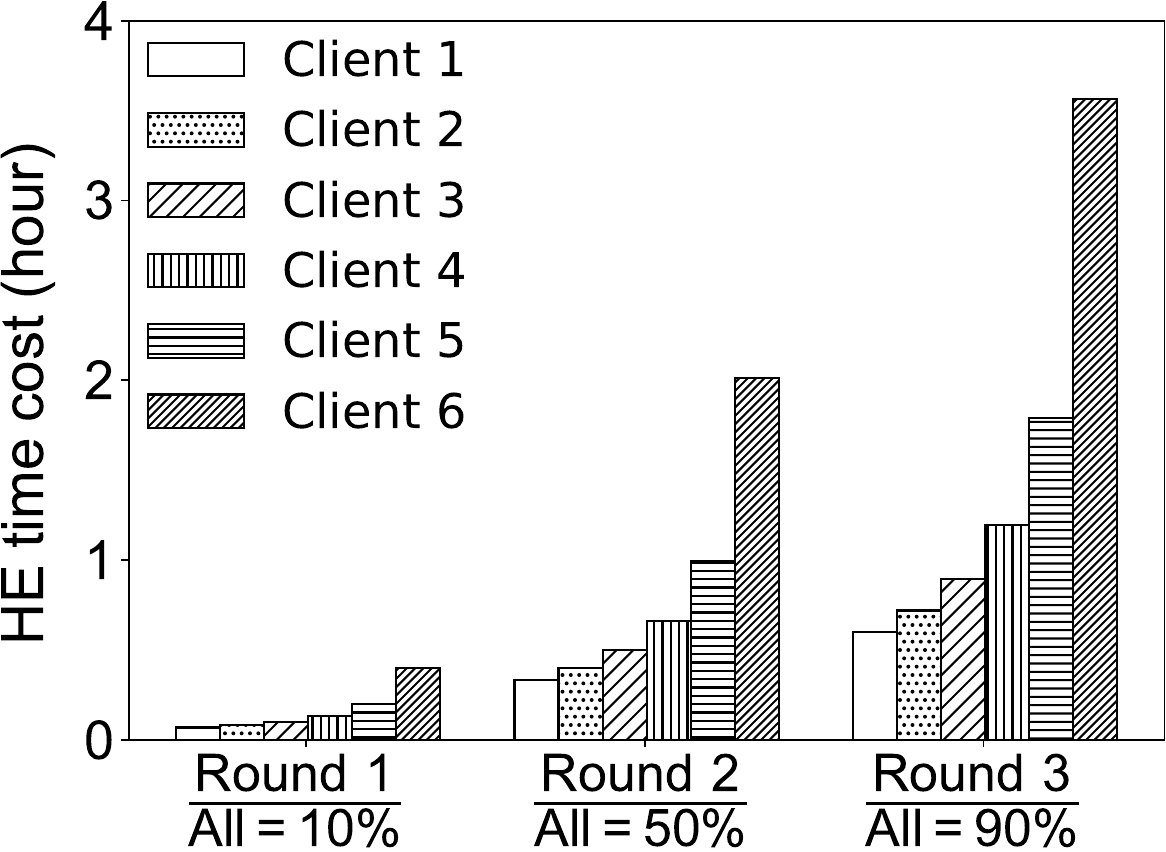}
    \caption{Imbalanced client HE time costs of existing Selective HE methods in system heterogeneity scenarios.}
    \label{sys_het_same_budget}
\end{figure}

Specifically, we set up 6 clients to conduct the FL training of an AlexNet 
on the \texttt{CIFAR10} dataset.
To emulate system heterogeneity, we allocate 6, 5, 4, 3, 2, 1 CPUs to the clients (denoted by \emph{Client 1-6}), respectively.
We conduct the FL training for 3 rounds, within which the model parameters are selectively encrypted by 
\emph{MaskCrypt} \cite{hu2024maskcrypt}
with the encryption budgets of 10\%, 50\% and 90\%, respectively.
Each round of FL training lasts for 3 iterations.
Then, for each client, we measure its mean HE time cost over all the iterations per round, which is illustrated in Figure \ref{sys_het_same_budget}.
We can see that when only 10\% of the model parameters are encrypted, the clients' HE time costs are not significantly different. 
This means that the overhead caused by FL training is not the major cause of straggling.
Along with the increase of the encryption budget, the imbalance of the clients' HE time costs deteriorates significantly.
For example, when the encryption budget increases from 10\% to 90\%, the HE time cost of Client 6 increases by \textgreater10$\times$, while the HE time cost of Client 1 only increases by \textless5$\times$. 
Thus, Clients 1-5 have to wait for Client 6 per model aggregation, which significantly delays the completion of FL training.
Therefore, we must maximally align the clients' HE time costs by adaptively tailoring the clients' encryption budgets according to their respective device capabilities. %for the sake of alleviating the straggler problem, 

\section{Performance Degradation of Existing Selective HE Methods in Statistical Heterogeneity Scenarios} \label{ana:stat_het}

\begin{figure}[t!]
    \centering
    \includegraphics[width=0.75\linewidth]{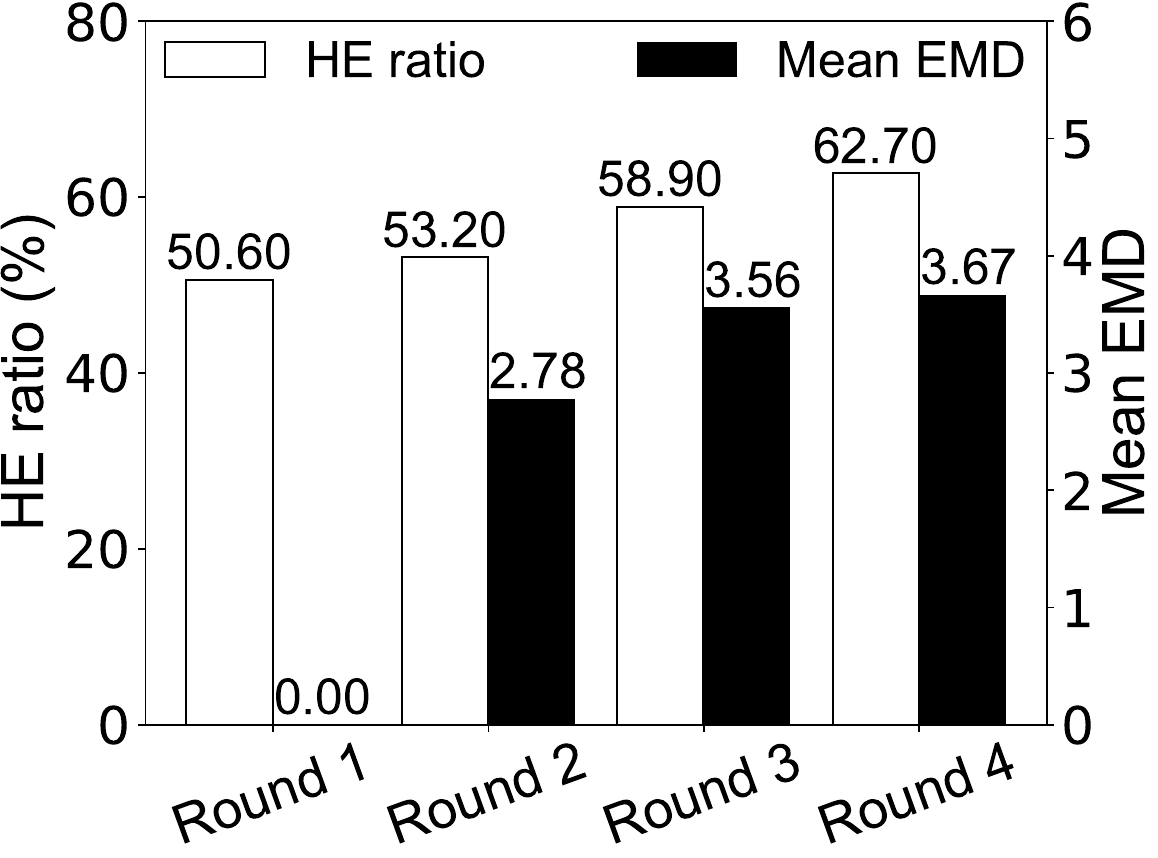}
    \caption{
HE ratios of existing Selective HE methods 
  under different degrees of statistical heterogeneity.
}
    \label{motivationC_union}
\end{figure}

To ensure the collaborative aggregation of encrypted model parameters, the existing Selective HE methods utilize the union of clients' selected model parameter subsets to represent the global selective encryption mask \cite{jin2023fedmlhe,hu2024maskcrypt}.
However, in statistical heterogeneity scenarios where clients have Non-IID training data, the encryption mask may be much larger than any clients' encryption subsets since the clients' data distributions 
may have little overlap.
In this case, the performance of HE overhead reduction of these methods may significantly degrade.
To confirm this conjecture, we vary the degree of statistical heterogeneity in the clients' training data and measure the corresponding HE ratios of model parameters.

Specifically, %we utilize the \texttt{CIFAR10} dataset for this experimental study.
to exclude the impact of system heterogeneity and data quantity difference, we set up 10 clients, each of which has 10\% data samples of \texttt{CIFAR10}, 1 CPU and a HE key size of 2048. %, which is the same as in \cite{zhang2020batchcrypt,zhang2021dubhe}.
We conduct the FL training for 4 rounds, with 3 iterations per round:
1) in the 1\textsuperscript{st} round, each client has all the 10 classes of data samples 
(i.e., strictly IID); 2) in the 2\textsuperscript{nd} and 3\textsuperscript{rd} rounds, each client has 5 and 2 classes of data samples, respectively; 3) in the 4\textsuperscript{th} round, each client only has 1 class of data samples (i.e., strictly Non-IID).
To visualize the statistical heterogeneity, we calculate the mean Earth Mover's Distance (EMD) between the training data of every two clients in each round, which measures the similarity between two statistical distributions \cite{zhang2021dubhe}.
The larger mean EMD that the clients result in, the more statistically heterogeneous they are in a round.
As illustrated with black bars in Figure \ref{motivationC_union}, the measured mean EMDs 
are: 0 (Round 1), 2.78 (Round 2), 3.56 (Round 3) and 3.67 (Round 4), respectively.
The white bars in Figure \ref{motivationC_union} illustrate the measured mean HE ratios of model parameters across varying degrees of statistical heterogeneity.
We can see that as the mean EMD increases, the HE ratio also significantly rises.
This indicates that the existing Selective HE methods become less cost-efficient in the presence of statistical heterogeneity.
To effectively minimize HE overhead through selective encryption, it is better to determine the HE mask over clients with IID data.
This motivates us to develop methods for measuring data distribution similarity, which can be utilized to cluster clients with IID data and determine cost-efficient HE masks for each cluster.

\section{Relation between Model Parameter Sensitivity and Data Distribution}\label{app:why_sen}
We know that model parameter sensitivity is intrinsically the same as gradients, both of which represent the change of loss values.
Since previous inversion attacks \cite{DBLP:conf/ccs/HitajAP17,dlg,geiping2020inverting} have demonstrated that the gradients transmitted during FL training can reveal clients' data distribution information, we can intuitively infer that the clients' model parameter sensitivity
is also consistent with their data distribution.
To confirm this, we continue to analyze the relationship between model parameter sensitivity and data distribution. %in statistical heterogeneity scenarios.

\begin{figure}[t!]
    \centering
    \includegraphics[width=0.75\linewidth]{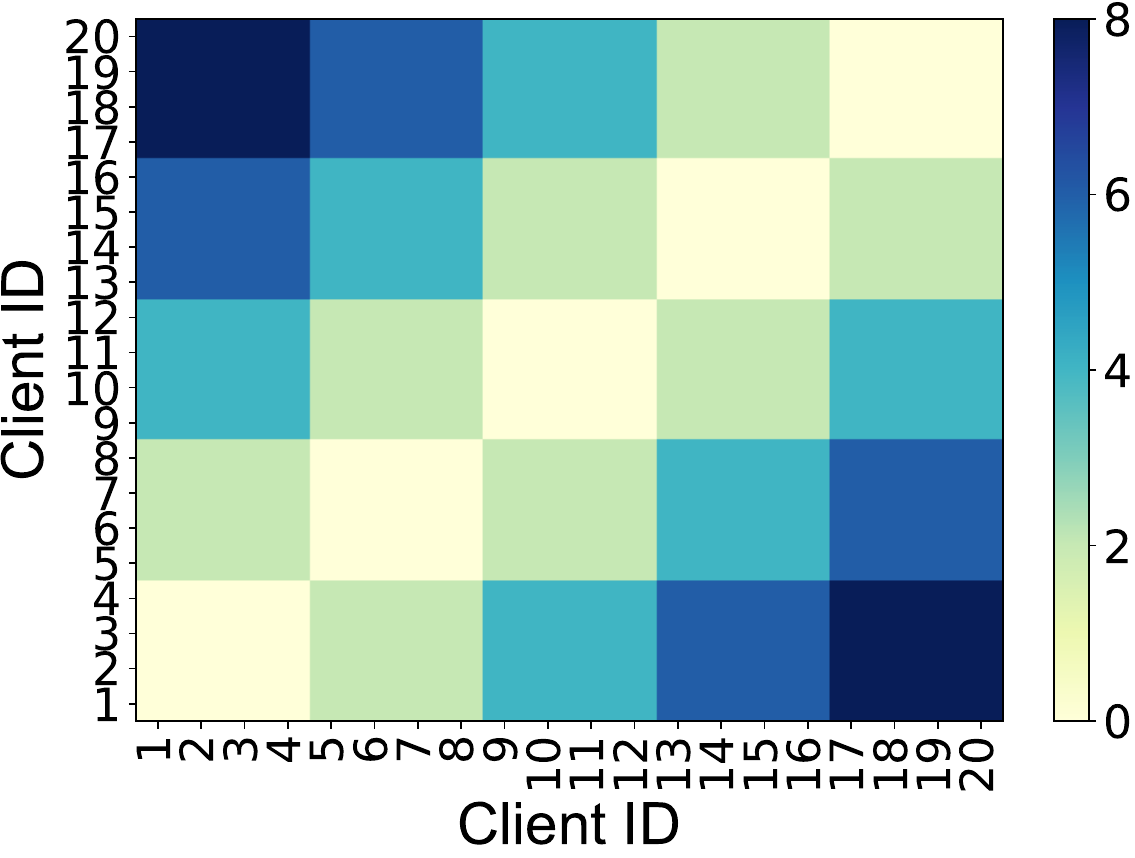}
    \caption{Heat map matrix of pairwise statistical heterogeneity in the training data of 20 clients.}
    \label{iid_non_iid}
\end{figure}

\begin{figure}[t!]
    \centering
    \includegraphics[width=0.75\linewidth]{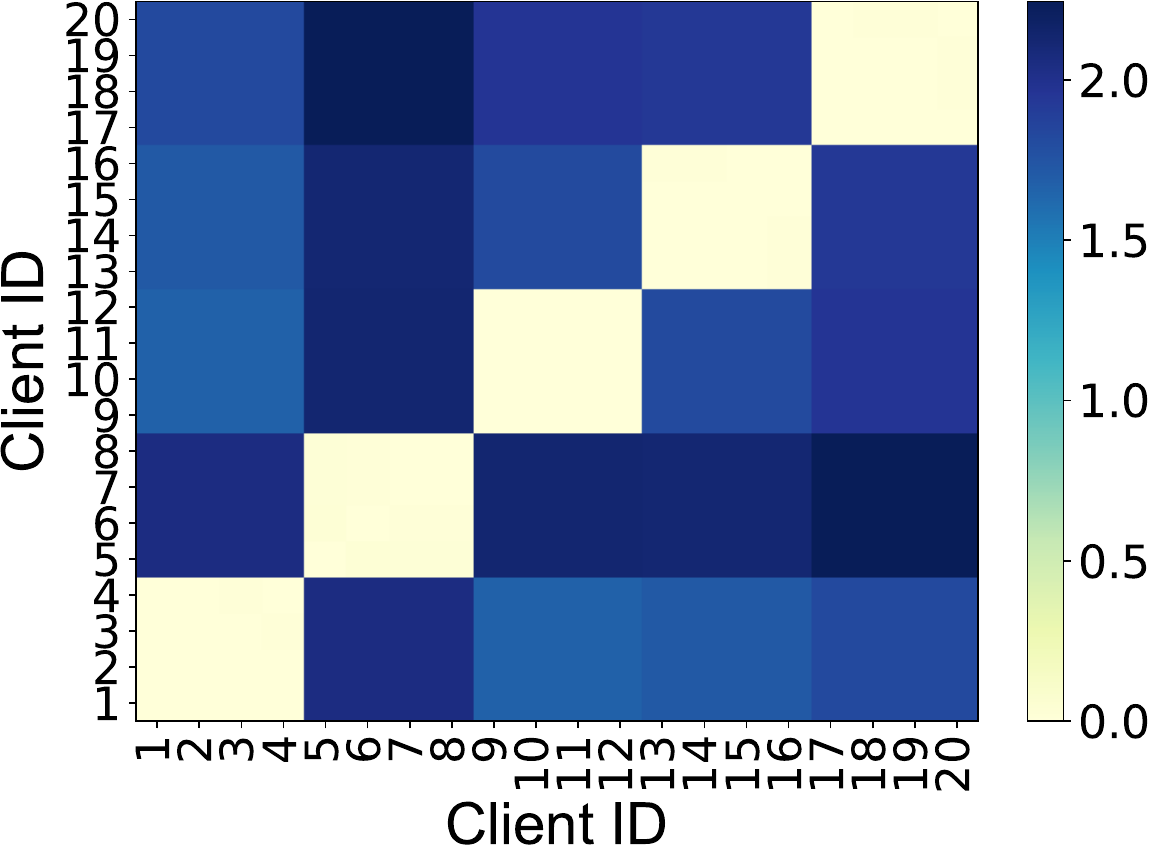}
    \caption{Heat map matrix of pairwise model parameter sensitivity similarity between the 20 clients.}
    \label{motivationC_iid_non_iid}
\end{figure}

Specifically, %we utilize the \texttt{CIFAR10} dataset for this experimental study.
we first set up 20 clients, each of which has 1 CPU and a HE key size of 2048. %, which is the same as in \cite{zhang2020batchcrypt,zhang2021dubhe}.
Then, we classify the clients into 5 categories with 4 clients per category, and ensure that 
every 2 classes of \texttt{CIFAR10} data samples are equally split to every 4 clients within the same category. %as the training data. %with the same data size.
Figure \ref{iid_non_iid} shows the heat map matrix of the pairwise EMD between every two clients.
The darker the color of the square corresponding to two clients, the more statistically heterogeneous they are.
As expected, the clients are clearly split into 5 categories, within which the clients' data distributions are IID.
Subsequently, we conduct the training of an AlexNet \cite{krizhevsky2012imagenet} for one iteration on each client, and obtain the gradients $\nabla_{\mathbf{W}}\mathcal{L}(\mathbf{W})$ and model parameter values $\mathbf{w}$.
Next, we calculate the Euclidean distance \cite{danielsson1980euclidean} between the sensitivity vectors of every two clients as their similarity, which is also illustrated as a heat map matrix in Figure \ref{motivationC_iid_non_iid}.
We can see that although Figure \ref{motivationC_iid_non_iid} does not completely follow Figure \ref{iid_non_iid}, the pairwise similarities between the sensitivity vectors of the clients within the same category (i.e., antidiagonal of Figure \ref{motivationC_iid_non_iid}) are highly consistent with those in Figure \ref{iid_non_iid}.
This means that model parameter sensitivity can provide an alternative for data distribution representation and similarity measurement.

\section{Selection of $B$ and $C$}
\label{app:selection_BC}

The values of $B$ and $C$ are scenario-dependent and must balance security, computational efficiency, and system heterogeneity. To guide their selection, we propose an empirical methodology based on the following objectives:

\begin{enumerate}[
fullwidth,
itemindent=1em,
label=(\arabic*),
topsep=0pt,
]

    \item Ensure the model parameter selection optimization problem has a valid solution under system heterogeneity.
    
    \item Reduce Mutual Information (MI) between the original model parameters $\mathbf{W}$ and the ones after selective encryption $\mathbf{W}_{-\mathbf{w}}$, thereby limiting privacy leakage. 
    
\end{enumerate}

\begin{figure}[t!]
    \centering
    \includegraphics[width=0.75\linewidth]{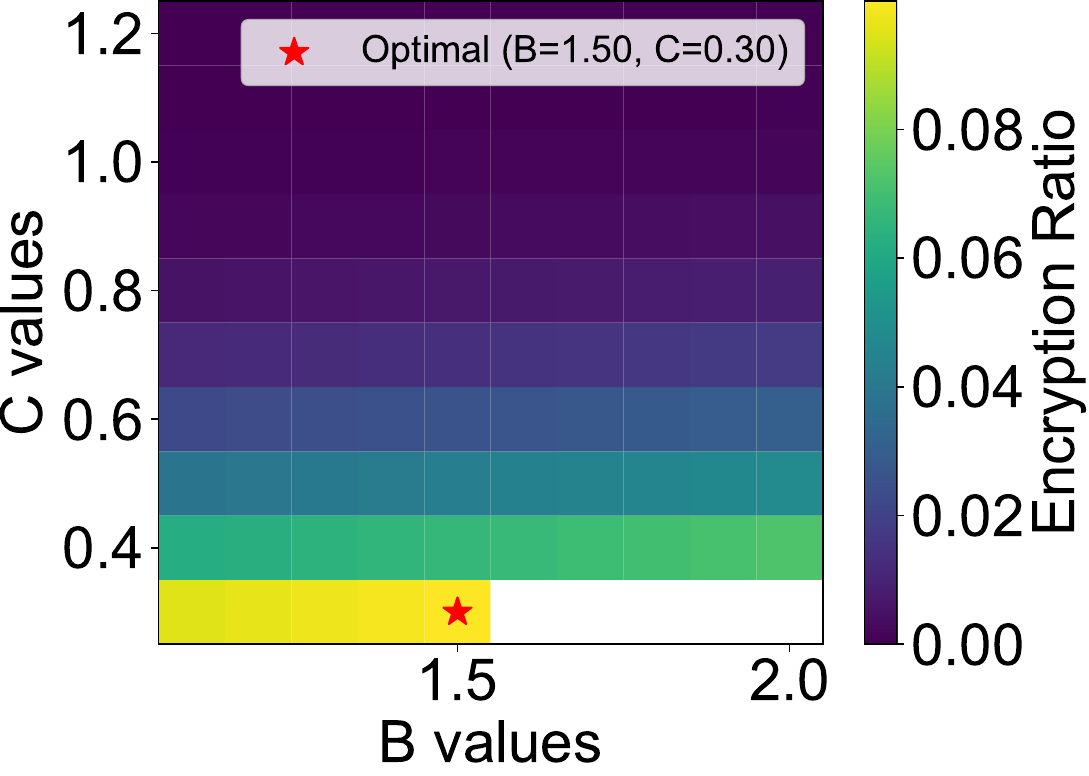}
    \caption{Encryption ratios under different $B$ and $C$ values.}
    \label{fig:BC_encryption_ratio_heatmap}
\end{figure}

\begin{figure}[t!]
    \centering
    \includegraphics[width=0.75\linewidth]{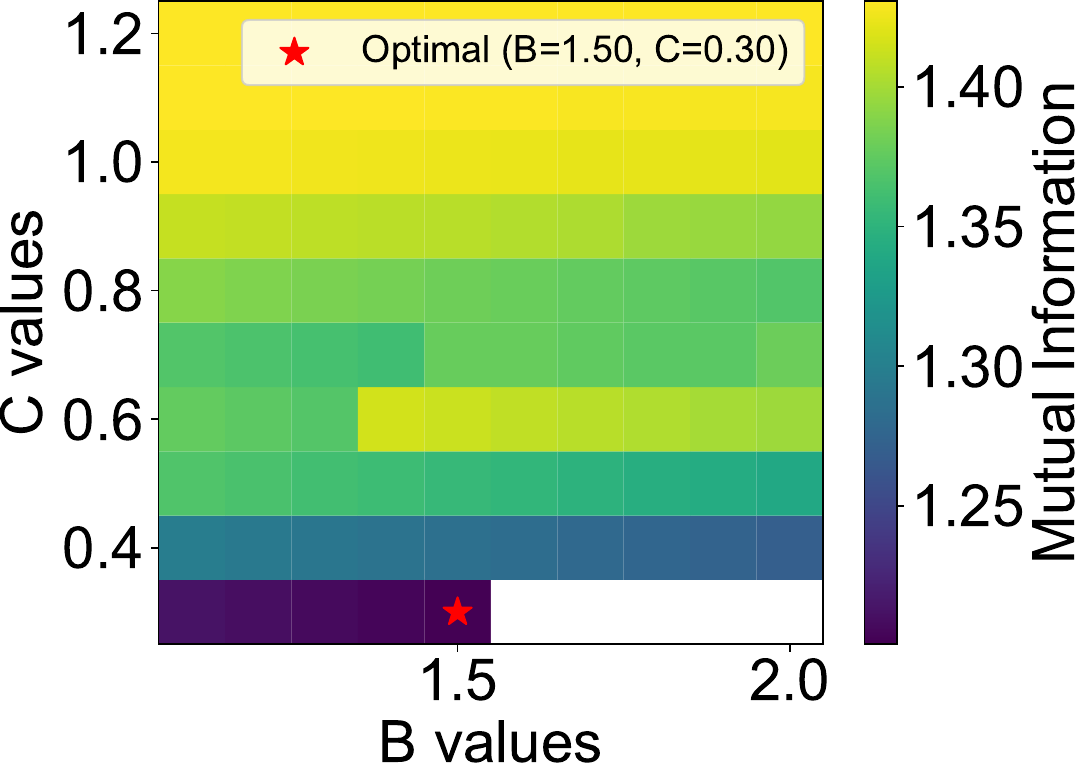}
    \caption{Mutual information under different $B$ and $C$ values.}
    \label{fig:BC_mutual_information_heatmap}
\end{figure}

Figures \ref{fig:BC_encryption_ratio_heatmap} and \ref{fig:BC_mutual_information_heatmap} illustrate the heat map matrix of encryption ratio and MI 
$I(\mathbf{W}, \mathbf{W}_{-\mathbf{w}})$ under different combinations of $B$ and $C$ values for a resource-constrained client (encryption budget $\alpha_i = 0.1$) on the \texttt{CIFAR-10} dataset, respectively. 
We can see that with the increase of encryption ratio, MI generally decreases.
The white squares in the heat map matrix represent the combinations of $B$ and $C$ values for which no feasible solutions can be obtained for the optimization problem.

Based on the empirical analysis results, we know that $(B=1.5, C=0.3)$ is the best combination that can achieve the maximum encryption ratio with the minimum MI.
Note that the selection of $B$ and $C$ values should adapt to the actual statistical and system heterogeneity condition of participating clients, which may shift or fluctuate during FL.
To improve the robustness of the encryption strategy, the selection of $B$ and $C$ values should tolerate certain variation margins (e.g., suboptimal combinations such as $(B=1.4, C=0.4)$).

\section{Clustering with DP-noised Sensitivity Vectors}
\label{app:sensitivity_noise}

Although no existing attack methods can directly reconstruct user data from sensitivity vectors, the inherent privacy risks cannot be entirely neglected.
To preemptively mitigate potential inference attacks, we adopt the noise injection methodology in DP-SGD \cite{Abadi_2016} for perturbing sensitivity vectors.

\begin{algorithm}[t!]
\caption{Privacy-Preserving Clustering of Clients with DP-noised Sensitivity Vectors.}
\label{alg:dp_clustering}
\SetKwFunction{Server}{Server}
\SetKwFunction{Client}{Client}
\SetKwFunction{Unbatching}{Unbatching}
\SetKwFunction{Batching}{Batching}
\SetKwInOut{Input}{Input}
\SetKwInOut{Output}{Output}

\Input{
Set of sensitivity vectors $\{\mathbf{\Gamma}_k\}$, 
noise scale $\sigma$,  
norm bound $G$
}
\Output{
Set of clusters $\{c_n\}$
}

\Fn{\Client}{
    Clip sensitivity vector: $\bar{\mathbf{\Gamma}}_k \leftarrow \mathbf{\Gamma}_k / \max(1, \|\mathbf{\Gamma}_k\|_2/G)$\;
    Add DP noise: $\tilde{\mathbf{\Gamma}}_k \leftarrow \bar{\mathbf{\Gamma}}_k + \mathcal{N}(0, \sigma^2 G^2 \mathbf{I})$\;
    \Return{\emph{$\tilde{\mathbf{\Gamma}}_k$}}
}

\Fn{\Server}{
    Initialize the set for storing sensitivity vectors $\{\tilde{\mathbf{\Gamma}}_k\}$\;
    \For{the $k$-th client}
    {
        $\{\tilde{\mathbf{\Gamma}}_k\} \leftarrow$ received DP-noised sensitivity vector $\tilde{\mathbf{\Gamma}}_k$\;
    }
    $\{c_n\} \leftarrow \text{AffinityPropagation}(\{\tilde{\mathbf{\Gamma}}_k\})$\;
    \Return{\emph{$\{c_n\}$}}
}
\end{algorithm}
\setlength{\textfloatsep}{0pt}% Remove \textfloatsep

\begin{figure}[t!]
\centering
\begin{subfigure}[t]{0.75\linewidth}
  \centering
  \includegraphics[width=\textwidth]{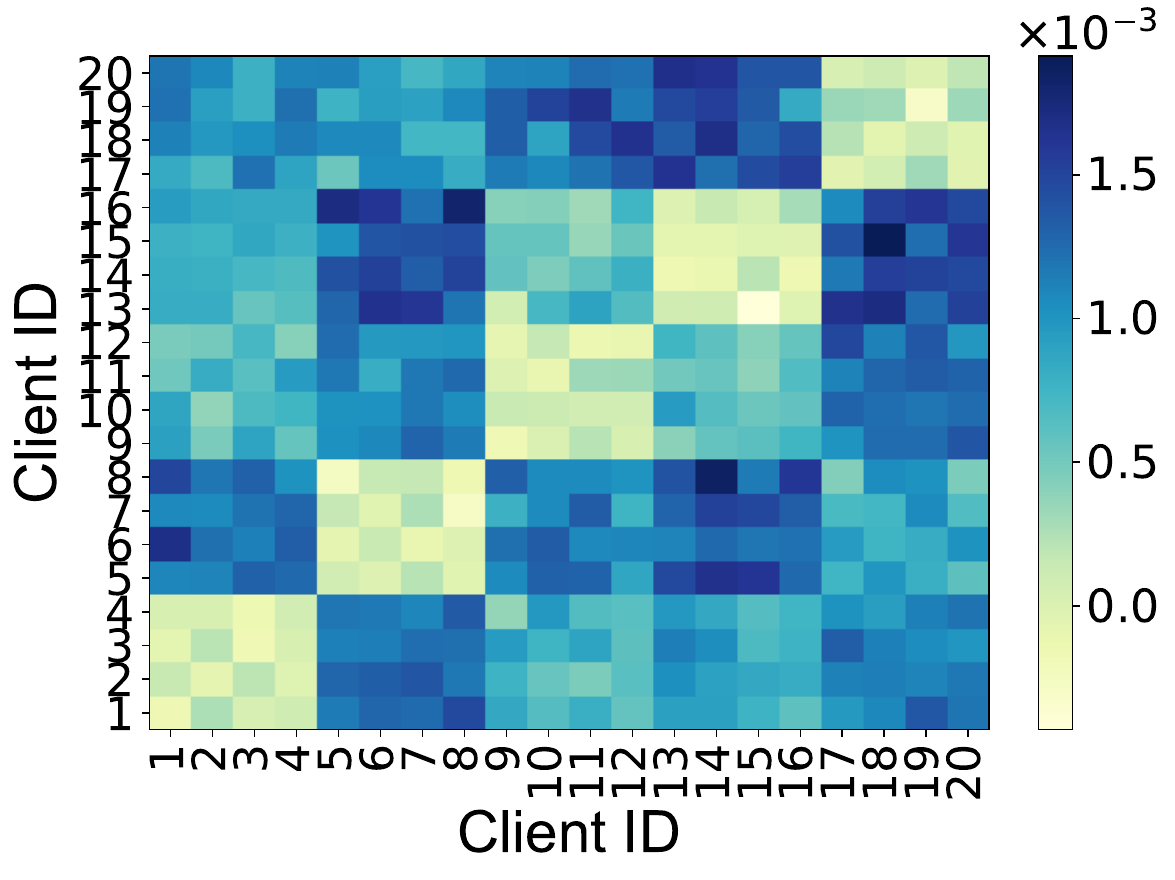}
  \caption{Higher privacy budget \\ ($\epsilon = 0.1$)
  }
  \label{fig:low_epsilon}
\end{subfigure}
\begin{subfigure}[t]{0.75\linewidth}
  \centering
  \includegraphics[width=\textwidth]{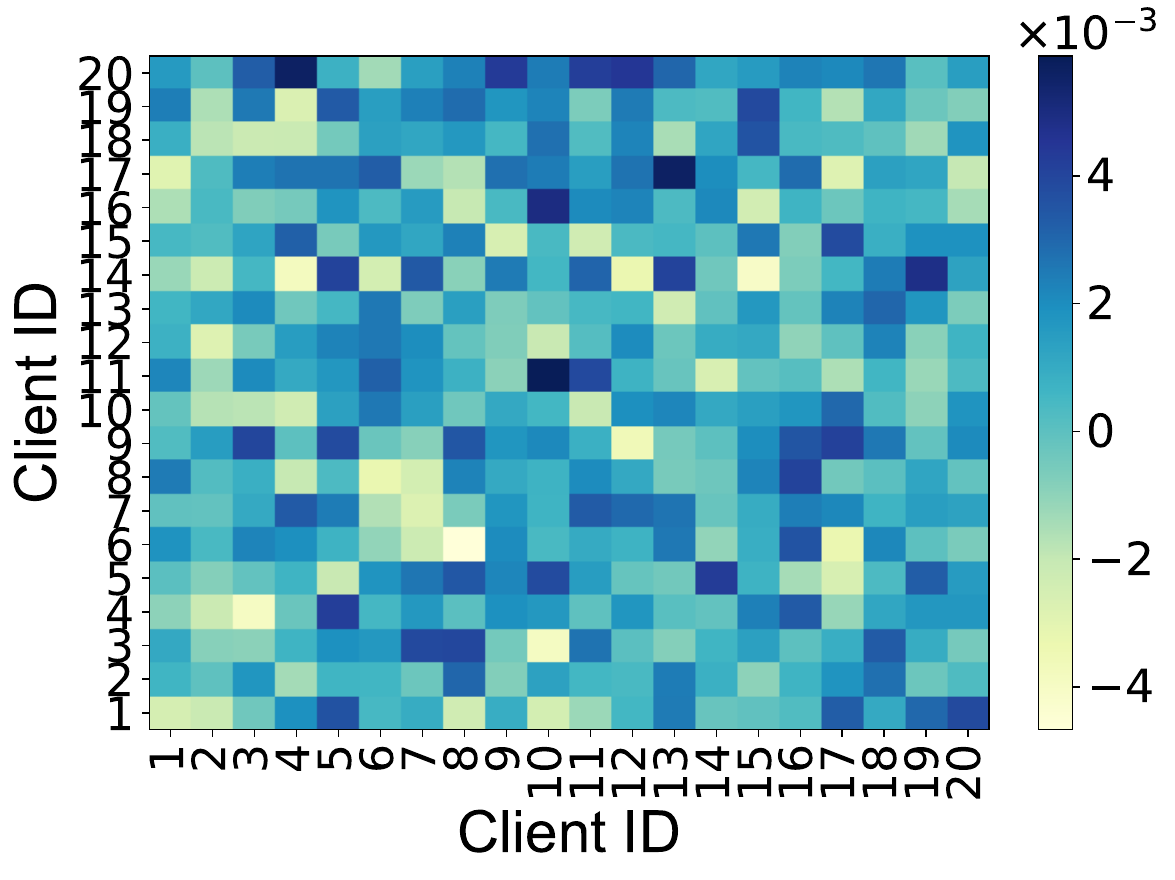}
  \caption{Lower privacy budget \\ ($\epsilon = 0.01$)}
  \label{fig:high_epsilon}
\end{subfigure}
\caption{Heat map matrix of pairwise similarity between the DP-noised sensitivity vectors of the 20 clients.}
\label{fig:epsilon_comparison}
\end{figure}

The noise magnitude $\sigma$ is calibrated to satisfy $(\epsilon, \delta)$-differential privacy through the Gaussian mechanism
$\sigma = \frac{\sqrt{2\ln(1.25/\delta)} }{\epsilon}$, where $\epsilon$ denotes the privacy budget. 
The lower the value of $\epsilon$, the stronger privacy protection that the sensitivity value will enjoy, but the greater error it will suffer.
$\delta$ represents the probabilistic relaxation parameter (fixed at $10^{-5}$), and $\sigma$ quantifies the standard deviation of the Gaussian noise injected into sensitivity vectors.

Algorithm \ref{alg:dp_clustering} presents the details of our proposed method for privacy-preserving clustering of clients based on their DP-noised sensitivity vectors.
The inputs include the set of clients' sensitivity vectors $\{\mathbf{\Gamma}_k\}$, the noise scale $\sigma$, and the norm bound $G$.
At line 2, each client performs $l_2$-norm clipping on its sensitivity vector, with the clipping bound $G$.
At line 3, each client adds the DP noise to its clipped sensitivity vector $\bar{\mathbf{\Gamma}}_k$.  
The noise follows the Gaussian distribution $\mathcal{N}(0, \sigma^2 G^2 \mathbf{I})$, with zero mean and covariance matrix $\sigma^2 G^2 \mathbf{I}$. %, where $G$ is the norm bound.
At lines 6-9, the FL server receives all clients' DP-noised sensitivity vectors, and applies the Affinity Propagation (AP) method to cluster the clients into their respective groups $c_n$, within which the clients' data distribution is IID to each other.

Following the experimental setup in Section \ref{ana:stat_het}, we further evaluate the effectiveness of client clustering based on DP-noised sensitivity vectors.
As shown in Figure \ref{fig:low_epsilon}, when $\epsilon=0.1$, although most pairwise similarity values are greatly perturbed, the similarity between the clients with IID data is still significantly higher than the others.
However, as shown in Figure \ref{fig:high_epsilon}, 
excessive noise injection over sensitivity values ($\epsilon = 0.01$) significantly hinders the effective measurement of data similarity.

\begin{figure}[t!]
    \centering
    \includegraphics[width=0.75\linewidth]{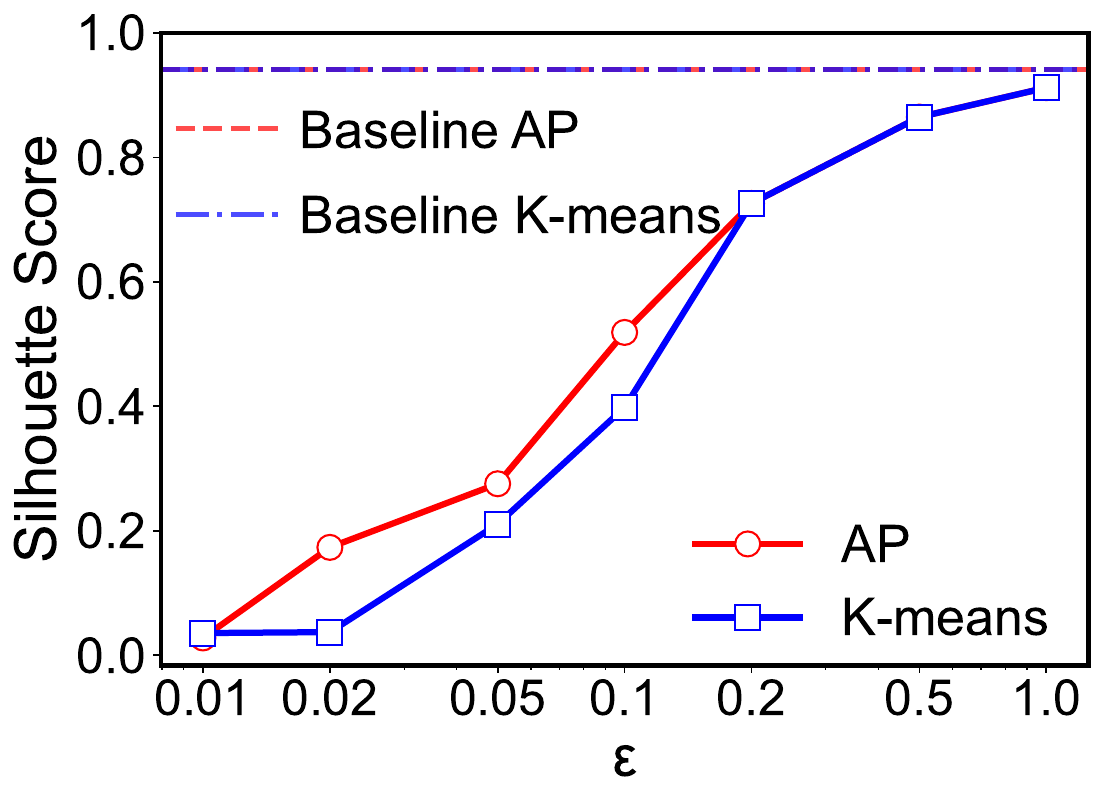}
    \caption{Impact of $\epsilon$ on Silhouette Score.}
    \label{fig:epsilon_vs_silhouette}
\end{figure}

\begin{figure}[t!]
    \centering
    \includegraphics[width=0.75\linewidth]{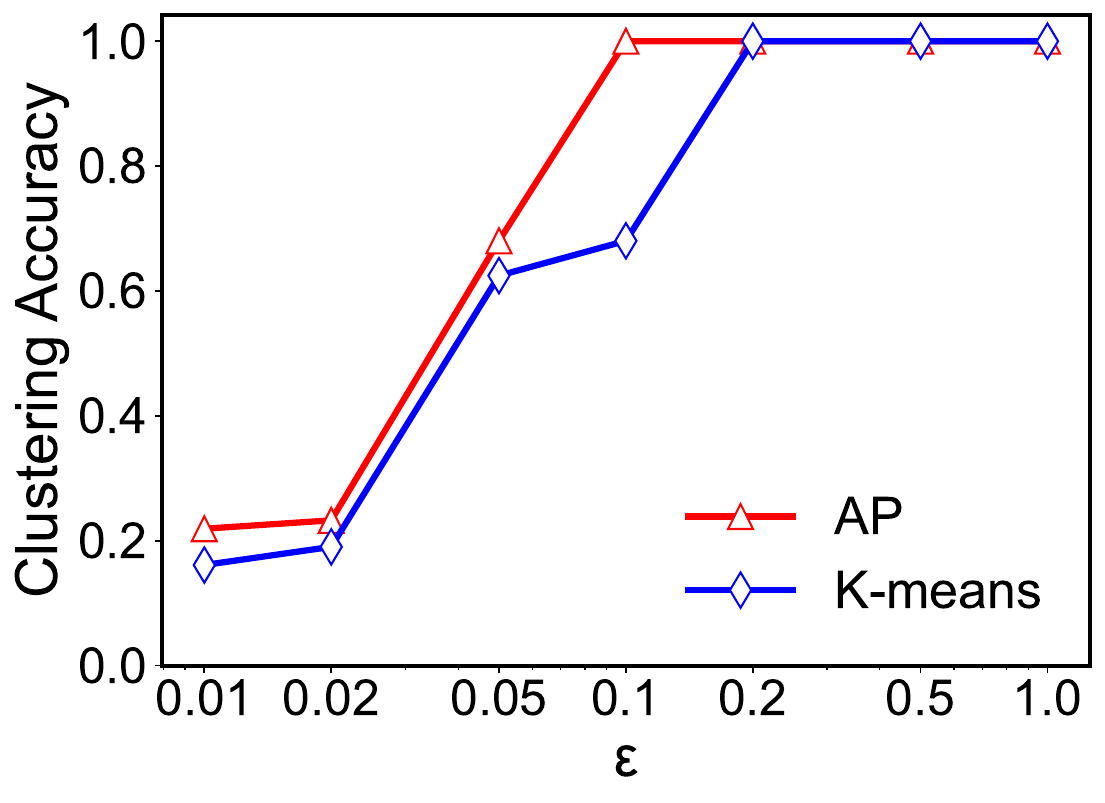}
    \caption{Impact of $\epsilon$ on Clustering Accuracy.}
    \label{fig:epsilon_vs_accuracy}
\end{figure}

Driven by these observations, we vary $\epsilon$ from 0.01 to 1.0 and apply the AP method and the K-means method to cluster the clients based on their DP-noised sensitivity vectors.
We use the Silhouette Score \cite{shahapure2020cluster,fan2024taking} and the Adjusted Rand Index (ARI) \cite{zhang2023enhancing} to evaluate clustering quality and clustering accuracy, respectively.
Specifically, the Silhouette Score ($S$) is an intrinsic metric for evaluating cluster quality based on cohesion and separation.
It is computed as: 
\begin{equation}
\label{equ:sscore}
S(i) = \frac{b(i) - a(i)}{\max(a(i), b(i))}, 
\end{equation}
where $a(i)$ is the mean distance between sample $i$ and all other data points in the same cluster, and $b(i)$ is the mean distance between sample $i$ and all the data points in its nearest cluster. 
$S(i)$ ranges from $-1$ to $1$, with higher scores indicating better clustering quality.

ARI reflects clustering accuracy by comparing the DP-noised clustering result with the baseline (i.e., ground truth clusters). 
It is computed as:

\begin{equation}
\small
\label{equ:ari}
\text{ARI}(C_n^p,C_n^g) = \frac{2(\beta_{00}\beta_{11} - \beta_{01}\beta_{10})}{(\beta_{00}+\beta_{01})(\beta_{01}+\beta_{11}) + (\beta_{00}+\beta_{10})(\beta_{10}+\beta_{11})}, 
\end{equation}
where $C_n^p$ is the predicted clustering result, $C_n^g$ is the ground truth clusters, $\beta_{11}$ is the number of pairs that are in the same cluster in both $C_n^p$ and $C_n^g$, $\beta_{00}$ is the number of pairs that are in different clusters in both $C_n^p$ and $C_n^g$, $\beta_{01}$ is the number of pairs that are in the same cluster in $C_n^p$ but in different clusters in $C_n^g$, and $\beta_{10}$ is the number of pairs that are in different clusters in $C_n^p$ but in the same cluster in $C_n^g$.
The ARI will be close to 0 if two clusters do not have any overlapped pair of data samples and exactly 1 if the clusters are the same.

Figures \ref{fig:epsilon_vs_silhouette} and \ref{fig:epsilon_vs_accuracy} illustrate the measured Silhouette Scores and accuracy of the clustering results under the two methods, respectively. 
\emph{Baseline AP} and \emph{Baseline K-means} represent the cases without DP-noise injection, \emph{AP} and \emph{K-means} represent the cases on DP-noised sensitivity vectors.
We can see that 
as the Silhouette Score is dependent on the distances between data samples, it is sensitive to subtle perturbation caused by DP noise.
In comparison, the clustering accuracy of both methods suffers no drop until $\epsilon$ decreases below 0.2.
This is mainly because that sensitivity vectors, as long as not excessively perturbed by noise, can effectively capture the underlying data distribution patterns, which validates the effectiveness of exploiting DP-noised sensitivity vectors for high-quality clustering of clients while preserving privacy.
What's more, we also notice that compared with the K-means method, which requires predefined number of clusters in advance, the AP method can automatically determine the number of clusters, and even tolerate the noise level at $\epsilon = 0.1$. 
This indicates that the AP method is more robust to DP noise perturbation than the K-means method and more effective for client clustering in scenarios where the server cannot predetermine the number of clusters.

\section{Effectiveness of Sensitivity-guided Selective HE}
\label{app:effectiveness_of_sense}

\begin{figure}[t!]
    \centering
    \includegraphics[width=0.95\linewidth]{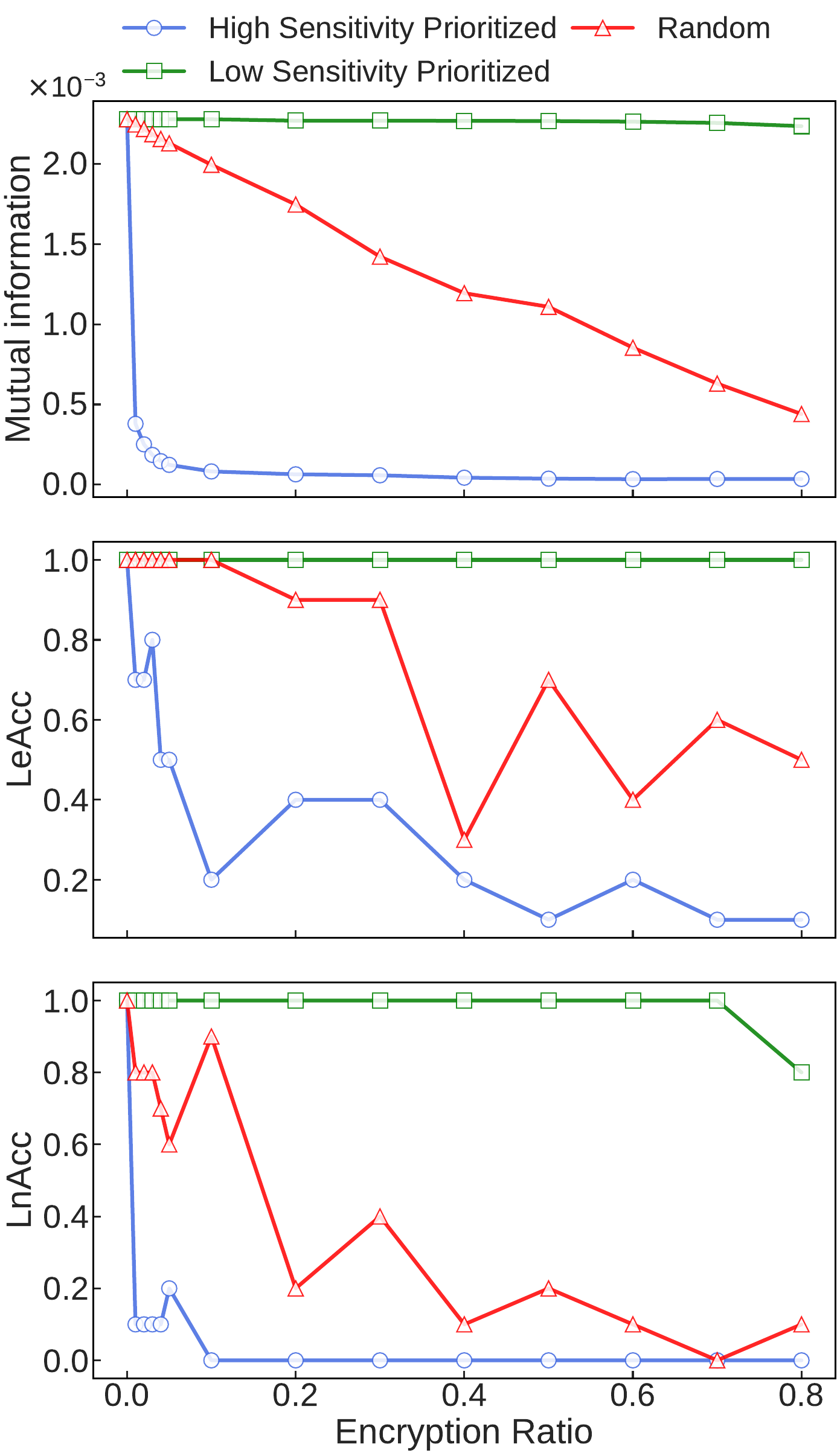}
    \caption{Impact of Encryption Ratio and Strategy on Mutual Information and Attack Success Rate.}
    \label{fig:metrics_comparison}
\end{figure}

To validate the effectiveness of the sensitivity-guided selective HE strategy, we make a detailed analysis of MI and attack success rate under varying encryption ratios using three different encryption strategies:

\begin{itemize}[
fullwidth,
topsep=0pt,
      itemindent=1em,
]
    \item \textbf{High Sensitivity Prioritized.} Given a specified encryption ratio, we prioritize the selection of model parameters with the highest sensitivity for encryption.
    \item \textbf{Low Sensitivity Prioritized.} Given a specified encryption ratio, we prioritize the selection of model parameters with the lowest sensitivity for encryption.
    \item \textbf{Random.} Given a specified encryption ratio, we randomly select model parameters for encryption.
\end{itemize}

Specifically, we launch the iLRG attack under varying encryption ratios ranging from 0 to 0.8, and measure the changes of MI, Label existence Accuracy (LeAcc) and Label number Accuracy (LnAcc). %as the encryption ratio ranged from 0 to 0.8.
The experiments are conducted on ResNet-50 with a batch size of 30, and the \texttt{MNIST} dataset. 
This set of experimental settings is vulnerable to iLRG attacks 
on plaintext, thereby excluding the influence of settings other than the encryption strategies. 

Figure \ref{fig:metrics_comparison} illustrates the measured results of each strategy.
We can see that along with the increasing of encryption ratio, the MI of \textbf{Random} almost decreases linearly.
While for the case of \textbf{Low Sensitivity Prioritized}, the MI almost remains unchanged, and for the case of \textbf{High Sensitivity Prioritized}, the MI sharply drops to almost 0 when the encryption ratio is above 0.1.
Correspondingly, LeAcc and LnAcc follow similar change patterns.
These results confirm the reliability of employing MI to quantify the privacy leakage caused by the selective encryption of model parameters.
What's more, we can also notice that \textbf{Low Sensitivity Prioritized} and \textbf{High Sensitivity Prioritized} demonstrate stark contrast in privacy protection in terms of LeAcc and LnAcc. 
Although \textbf{Random} can finally reach a high privacy protection level, it consumes significantly more encryption overhead than \textbf{High Sensitivity Prioritized}.

In summary, these observations strongly advocate the effectiveness and necessity of applying sensitivity-guided selective HE strategy, which validates the theoretical foundation of our proposed framework that adaptively balances security and HE overhead for each client in cross-device FL.

\section{HE Key Management and Distribution}
\label{app:key_management}

In the standard threat model, the server is honest-but-curious, and clients are honest. However, a more realistic threat model considers the presence of malicious clients who might collude with the server or other clients. 
If all clients share a single private key, a malicious client could intercept and decrypt the model updates from any other client, completely compromising their privacy.

To address this expanded threat, we propose a framework that decouples the aggregation server from a trusted \textbf{Decryption and Key Management Server (DKMS)}. In our framework, the DKMS assumes the critical role of generating and distributing public/private key pairs, as well as decrypting the final aggregated model. This design does not introduce additional physical infrastructure; instead, it formally delegates key management and decryption operations to a trusted third party - a function that is frequently implied but rarely explicitly defined in most homomorphic encryption-based federated learning systems. Maintaining the private key either at the aggregation server or with participating clients would compromise security, as possession of the private key would enable decryption of all sensitive communications. The modified complete algorithm pseudocode can be referred to in Algorithm \ref{alg:full_framework}.

This decoupled architecture offers two significant advantages:
\begin{enumerate}
    \item \textbf{Enhanced Security:} By isolating the private key on a dedicated DKMS, we prevent the aggregation server and any potentially malicious clients from decrypting individual client updates. The aggregation server only performs homomorphic additions on ciphertexts, and clients only receive the final, decrypted global model from the DKMS.
    \item \textbf{Improved Efficiency in Heterogeneous Environments:} In our system, clients only perform encryption. The final aggregated model is decrypted once by the DKMS and then distributed to all clients. This eliminates the need for each client to perform decryption locally, which can be a significant bottleneck, especially for resource-constrained devices. Since the final aggregated model requires a union of all clients' encryption masks for decryption, the decryption time would be dictated by the slowest device, exacerbating the straggler problem. Centralizing decryption at the DKMS effectively mitigates this issue.
\end{enumerate}

\begin{algorithm}[t!]
\caption{Workflow of \emph{SenseCrypt} with Dual-Server Architecture.}
\label{alg:full_framework}
\small
\SetKwFunction{DKMS}{DKMS}
\SetKwFunction{AggregationServer}{AggregationServer}
\SetKwFunction{Client}{Client}
\SetKwInOut{Input}{Input}
\SetKwInOut{Output}{Output}

\Input{
	Set of clients $\{1, \dots, K\}$,  parameters $B, C, \eta_{\text{MI}}$\;
}
\Output{Converged global model $\mathbf{W}^{\text{final}}$\;}

\BlankLine

\textbf{Phase 1: Initialization}\;
\DKMS{}:
    \parbox[t]{7.5cm}{
        Generate HE key pair: $(\text{pk}, \text{sk}) \leftarrow \text{Paillier.keygen()}$\;
        Distribute $\text{pk}$ to all clients\;
    }
\For{each client $i=1, \dots, K$ \textbf{\emph{in parallel}}}{
    \Client{$i$}:
    \parbox[t]{7.5cm}{
        Local training to get initial $\mathbf{W}_i^1$ and $\nabla_{\mathbf{W}}\mathcal{L}(\mathbf{W}_i^1)$\;
        Calculate sensitivity vector: $\mathbf{\Gamma}_i \leftarrow \lvert (\mathbf{W}_i^1)^\top \nabla_{\mathbf{W}}\mathcal{L}(\mathbf{W}_i^1) \rvert$\;
        Send $\mathbf{\Gamma}_i$ and $(r_i, v_i)$ to \AggregationServer{}\;
    }
}
\AggregationServer{}:
    \parbox[t]{7.5cm}{
        Receive $\mathbf{\Gamma}_i$ and $(r_i, v_i)$ from all clients\;
        \For{each client $i=1, \dots, K$}{
            Calculate encryption budget: \\ 
            $$\alpha_i \leftarrow \frac{\min \{ \overline{r}_i, \overline{v}_i \}}{\max \{ \lvert \min \{ \overline{r}_j, \overline{v}_j \} \rvert \}_{j=1}^K} \;$$
            Send $\alpha_i$ back to client $i$\;
        }
        Cluster clients based on sensitivity similarity: \\
        	\parbox[t]{7.5cm}{ $\{G_{\text{IID}}\} \leftarrow \text{AffinityPropagation}(\{\mathbf{\Gamma}_1, \dots, \mathbf{\Gamma}_K\})$\;}
    }

\BlankLine

\textbf{Phase 2: Federated Training Loop}\;
\For{each iteration $e=1, 2, \dots$}{
    \For{each group $G_{\emph{IID}} \in \{G_{\text{IID}}\}$ \textbf{\emph{in parallel}}}{
    \For{each client $i \in G_{\emph{IID}}$ \textbf{\emph{in parallel}}}{
    \Client{$i$}:
    \parbox[t]{7cm}{
        Receive global model $\mathbf{W}^{e}$ \;
        Update local model: $\mathbf{W}_i^e \leftarrow$ local training on $\mathbf{W}^{e}$\;
        Update sensitivity vector: $\mathbf{\Gamma}_i \leftarrow \lvert (\mathbf{W}_i^e)^\top \nabla_{\mathbf{W}}\mathcal{L}(\mathbf{W}_i^e) \rvert$\;
        Solve for optimal mask $\mathbf{X}_i = [x_k^i] \in \{0,1\}^{N^{\mathbf{w}}}$ by: \\
        $\min \quad [\sum_{k=1}^{N^{\mathbf{w}}} x_k^i, -\sum_{k=1}^{N^{\mathbf{w}}} x_k^i \gamma_k^i]^\top$ \\
        \textbf{s.t.}:\\
        1. $\sum_{k=1}^{N^{\mathbf{w}}} x_k^i \leqslant \lfloor \alpha_i N^{\mathbf{w}} \rfloor$ \quad\textit{(Budget Constraint)}\\
        2. $\frac{\sum x_k^i \gamma_k^i}{\sum \gamma_k^i} \geqslant 1 - C e^{- B \alpha_i}$ \quad\textit{(Security Constraint)}\\
        3. $I(\mathbf{W}; (1 - \mathbf{X}_i) \odot \mathbf{W}) \leq \eta_{\text{MI}}$ \quad\textit{(MI Constraint)}\\
        Selectively encrypt model: $$\mathbf{W}_i^{e,*} \leftarrow \text{Encrypt}(\mathbf{W}_i^e, \mathbf{X}_i, \text{pk})\; $$
        Send $\mathbf{W}_i^{e,*}$ and  $\mathbf{X}_i$ to \AggregationServer{}\;   
            }
        }
      \AggregationServer{}: \parbox[t]{7.5cm}{
        Receive $\mathbf{W}_i^{e,*}$ and $\mathbf{X}_i$ from all clients in $G_{\text{IID}}$\;
        Aggregate encrypted models: $$\mathbf{W}^{e+1,*} \leftarrow \sum_{i \in G_{\text{IID}}} \frac{n_i}{N_{G}} \mathbf{W}_i^{e,*}\; $$
        Create union mask for decryption: $\widehat{\mathbf{X}} \leftarrow \bigcup_{i \in G_{\text{IID}}} \mathbf{X}_i$\;
        Send $\mathbf{W}^{e+1,*}$ and  $\widehat{\mathbf{X}}$ to \DKMS{}\;
        }
            
    \DKMS{}:
       \parbox[t]{7.5cm}{
        Receive $\mathbf{W}^{e+1,*}$ and $\widehat{\mathbf{X}}$ from \AggregationServer{}\;
        Decrypt aggregated model: $$\mathbf{W}^{e+1} \leftarrow \text{Decrypt}(\mathbf{W}^{e+1,*}, \widehat{\mathbf{X}}, \text{sk})\; $$
        Distribute plaintext global model $\mathbf{W}^{e+1}$ to all clients;
        }
    }
}
\end{algorithm}

\section{Proof of Collusion Resistance}
\label{app:collusion_resistance}

\textbf{Theorem 1:} If the aggregation server is honest-but-curious, and the server colludes with at most $|G_{IID}| - 2$ clients within a cluster $G_{IID}$, \emph{SenseCrypt} achieves model confidentiality for the honest clients in that cluster.

\textbf{Proof:} Let $C$ and $H$ denote the set of corrupted and honest clients respectively, within a specific cluster $G_{IID}$, where $|C| + |H| = |G_{IID}|$. Since the homomorphic encryption is IND-CPA secure, adversaries cannot infer an honest client's plaintext model $\mathbf{W}_i$ directly from its encrypted version $\mathbf{W}_i^*$.

However, adversaries can access the final decrypted aggregation result $\mathbf{W}^{e+1}$ for their cluster. Following the aggregation protocol, the server and the colluding clients in $C$ can compute the aggregated model of the honest clients by subtracting their own contributions:
$$ \sum_{i \in H} p_i \mathbf{W}_i = \mathbf{W}^{e+1} - \sum_{j \in C} p_j \mathbf{W}_j $$
If there is only one honest client (i.e., $|H|=1$), the adversary can recover the full model update $\mathbf{W}_i$ of that client. However, as long as there are at least two honest clients (i.e., $|H| \ge 2$), adversaries can only obtain the aggregated sum of their models, but cannot disentangle this sum to recover the individual model update $\mathbf{W}_i$ of any specific honest client.

Therefore, given the server colludes with at most $|G_{IID}| - 2$ clients, ensuring $|H| \ge 2$, \emph{SenseCrypt} protects the model confidentiality of individual honest clients within the cluster.

\section{Choice of Encryption Method: Paillier vs. CKKS}
\label{app:encryption_choice}

While both Paillier and schemes like CKKS (Cheon-Kim-Kim-Song) are popular homomorphic encryption solutions, Paillier is better suited for the \emph{SenseCrypt} framework. The primary reason lies in the incompatibility of our selective encryption strategy with the batching (or packing) technique that makes CKKS highly efficient.

In \emph{SenseCrypt}, the decision to encrypt a parameter is based on its individual sensitivity score. This means that sensitive and non-sensitive parameters can be adjacent to one another in the model's structure. Furthermore, each client generates a unique encryption mask tailored to its data and system capabilities.

CKKS achieves its efficiency by packing multiple plaintext values (a vector) into a single ciphertext and performing SIMD (Single Instruction, Multiple Data) operations. This requires the data to be arranged in a specific order before encryption. In our case, to use batching, clients would need to reorder their model parameters, separating the ones to be encrypted from the ones to be sent in plaintext. Since each client has a different mask, their reordering would be different, and the server would be unable to perform meaningful aggregation because the parameter positions would not align.

\begin{table}[t!]
\centering
\resizebox{\columnwidth}{!}{%
\begin{tabular}{@{}cccc@{}}
\toprule
\textbf{Quantity} & \textbf{Scheme} & \textbf{Encryption Time (s)} & \textbf{Ciphertext Size (bytes)} \\ \midrule
\multirow{2}{*}{1}    & Paillier & 0.3748                       & 768                            \\
                      & CKKS     & 0.0073                       & 333324 $\times$ 1                           \\ \midrule
\multirow{2}{*}{10}   & Paillier & 3.7613                       & 7680                            \\
                      & CKKS     & 0.0076                       & 333570  $\times$ 10                        \\ \midrule
\multirow{2}{*}{100}  & Paillier & 40.7887                      & 76800                           \\
                      & CKKS     & 0.0120                       & 334488  $\times$ 100                        \\ \midrule
\multirow{2}{*}{1000} & Paillier & 396.1321                     & 767990                          \\
                      & CKKS     & 0.0079                       & 334366  $\times$ 1000                        \\ \midrule
\multirow{2}{*}{10000} & Paillier & 3925.9769                     & 7679907                          \\
                      & CKKS     & 0.0261                       & 1002800  $\times$ 10000                        \\ \bottomrule
\end{tabular}%
}
\caption{Comparison of Encryption Performance between Paillier and CKKS Schemes.}
\label{tab:he_performance_comparison_resized}
\end{table}

Therefore, we cannot leverage the batching capabilities of CKKS. As shown in Table \ref{tab:he_performance_comparison_resized}, when encrypting a single parameter without batching, Paillier is far more practical. Although CKKS is faster for the encryption operation itself, its resulting ciphertext is designed to hold a large vector and is therefore massive, regardless of the number of encrypted values. Specifically, encrypting a single parameter with CKKS would generate a ciphertext of 333324 bytes, which is over \textbf{434 times} larger than Paillier's 768 byte ciphertext. This dramatic increase in data size would lead to an untenable communication overhead. Given our parameter-wise selective strategy, Paillier offers a more efficient solution by minimizing this overhead.

\section{Time Consumption Analysis of Components}
\label{app:time_consumption}

To understand the computational overhead of each component in \emph{SenseCrypt}, we profiled the execution time for a single client during one training iteration on the \texttt{CIFAR10} dataset. The results are summarized in Table \ref{tab:time_consumption}.

As the table shows, the most time-intensive operations on the client side are encryption and solving the optimization problem. Local model training and decryption also contribute significantly to the overall latency. In contrast, the initial one-off cost of clustering on the server is negligible, and the sensitivity calculation on the client side is very fast. The server-side aggregation time is comparable to the client-side encryption time. These results underscore the importance of our approach to adaptively manage the encryption overhead to mitigate the straggler problem.

\begin{table}[t!]
\centering
\small
\begin{tabular}{l|c|c|c}
\hline
\textbf{Step} & \textbf{Min (s)} & \textbf{Max (s)} & \textbf{Avg (s)} \\
\hline
Local training & 7.1 & 7.7 & 7.4 \\
Sensitivity calculation & 0.03 & 0.15 & 0.1 \\
Optimization problem solving & 11.8 & 17.7 & 13.2 \\
Encryption & 79.7 & 103.8 & 87.3 \\
Decryption & 24.8 & 86.4 & 56.8 \\
\hline
FedAvg aggregation (server-side) & - & - & 87.9 \\
Clustering (server-side, one-off) & - & - & 0.002 \\
\hline
\end{tabular}
\caption{Average time consumption per iteration for a single client.}
\label{tab:time_consumption}
\end{table}

\section{Approximation of Parameter Sensitivity}
\label{app:sensitivity_approximation}

Our use of a first-order Taylor expansion is a well-established and reasonable method for approximating parameter importance. This technique is standard in machine learning, notably in classic network pruning algorithms. Here is the proof of the approximation expansion.
The sensitivity of a parameter subset $\mathbf{w}$ is defined as the change in loss when $\mathbf{w}$ is zeroed-out: $\mathbf{\Gamma}(\mathbf{w}) = \lvert \mathcal{L}(\mathbf{W}) - \mathcal{L}(\mathbf{W}_{-\mathbf{w}}) \rvert$. We can approximate the term $\mathcal{L}(\mathbf{W}_{-\mathbf{w}})$ by performing a Taylor series expansion of the loss function $\mathcal{L}$ around the point $\mathbf{W}$:
\[
\begin{split}
\mathcal{L}(\mathbf{W}_{-\mathbf{w}}) = \mathcal{L}(\mathbf{W}) 
    &+ (\mathbf{W}_{-\mathbf{w}} - \mathbf{W})^\top \nabla_{\mathbf{W}}\mathcal{L}(\mathbf{W}) \\
    &+ \frac{1}{2} \bigl((\mathbf{W}_{-\mathbf{w}} - \mathbf{W})^2\bigr)^\top \mathbf{H} (\mathbf{W}_{-\mathbf{w}} - \mathbf{W}) \\
    &+ \dotsb
\end{split}
\]
where $\mathbf{H}$ is the Hessian matrix of $\mathcal{L}$ evaluated at $\mathbf{W}$, and the ellipsis represents higher-order terms.

The difference vector $(\mathbf{W}_{-\mathbf{w}} - \mathbf{W})$ is a vector that is equal to $-\mathbf{w}$ at the positions corresponding to the parameters in the subset $\mathbf{w}$, and is zero everywhere else. The first-order approximation is made by truncating the series after the linear term, assuming that the quadratic and higher-order terms are negligible. This is a common assumption when the change (in this case, zeroing out $\mathbf{w}$) is relatively small. This truncation leads to the approximation:
\[
\begin{split}
\mathcal{L}(\mathbf{W}_{-\mathbf{w}}) - \mathcal{L}(\mathbf{W}) 
    &\approx (\mathbf{W}_{-\mathbf{w}} - \mathbf{W})^\top \nabla_{\mathbf{W}}\mathcal{L}(\mathbf{W}) \\
    &= -\mathbf{w}^\top \nabla_{\mathbf{W}}\mathcal{L}(\mathbf{W})
\end{split}
\]

Rearranging the terms, we get the change in loss:
$$ \mathcal{L}(\mathbf{W}) - \mathcal{L}(\mathbf{W}_{-\mathbf{w}}) \approx \mathbf{w}^\top \nabla_{\mathbf{W}}\mathcal{L}(\mathbf{W}) $$

Substituting this approximation back into the sensitivity definition, we obtain the final, computationally tractable formula:
$$ \mathbf{\Gamma}(\mathbf{w}) \approx \lvert \mathbf{w}^\top \nabla_{\mathbf{W}}\mathcal{L}(\mathbf{W}) \rvert $$

\section{Related Work} \label{sec:related}

\textbf{Statistical \& system heterogeneity in FL.}
Cho \textit{et al.} \cite{cho2022flame} proposed to personalize model training for countering statistical heterogeneity in cross-device FL.
Liu \textit{et al.} \cite{liu2021pfa} proposed to utilize the sparsity of CNNs for privacy-preserving similarity measurement of clients' data.
To mitigate system heterogeneity issues, Bonawitz \textit{et al.} \cite{bonawitz2019towards} designed a FL system that filters clients by their response speed in 
FL training.
Chai \emph{et al.} \cite{chai2020tifl} proposed a tier-based FL system that
selects clients by training performance
to mitigate 
system heterogeneity problems.
Zhou \textit{et al.} \cite{zhou2023reinforcement} proposed to cluster clients by device
capabilities, and utilize reinforcement learning to mitigate statistical heterogeneity issues.
Jiang \emph{et al.} \cite{jiang2023heterogeneity} proposed to
utilize adaptive client selection and gradient compression for addressing the straggler problem.
However, these methods mostly rely on additional components to measure data similarity or coordinate FL training, which creates extra burden for FL framework adaptation.

\textbf{Overhead-optimized HE for FL.}
To make HE more lightweight, Jiang \emph{et al.} \cite{jiang2021flashe} proposed to drop the asymmetric-key design to meet the minimum requirements of security and functionality.
Hao \emph{et al.} \cite{8761267} proposed to
combine lightweight symmetric additive HE 
with DP technique to improve security protection.
However, such methods generally suffer from degraded security since symmetric encryption cannot provide the same level of security as public-key encryption.

Another alternative
is to encrypt multiple model parameters in one go.
Zhang \emph{et al.} \cite{zhang2020batchcrypt} proposed \textit{BatchCrypt}, which utilizes batch quantization and encryption of model parameters to reduce HE overhead.
Chen \emph{et al.} \cite{chen2021secureboost} further migrated \textit{BatchCrypt} to the establishment of gradient boosting decision trees for vertical FL.
Xu \emph{et al.} \cite{xu2021efficient} indicated that the gradient quantization in \textit{BatchCrypt} may cause frequent overflow errors due to the addition of negative numbers represented in two's complement, and result in accuracy loss.
To fix this issue, Zheng \emph{et al.} \cite{zheng2022aggregation} modified \textit{BatchCrypt} via down-scaling and identification of negative model parameter values.
Han \emph{et al.} \cite{hanyan} proposed an adaptive accuracy-lossless batch HE method that shifts parameters to non-negative values for the prevention of overflow errors.
Nevertheless, these methods 
cannot easily fit in mainstream FL frameworks without specific adaptation.

Inspired by the model pruning methods that remove redundant model parameters by sensitivity \cite{DBLP:conf/iclr/MolchanovTKAK17,DBLP:journals/corr/abs-1801-05787,NEURIPS2019_f34185c4,Molchanov_2019_CVPR,DBLP:conf/iclr/LubanaD21}, several methods proposed to selectively encrypt relatively more sensitive model parameters to reduce HE overhead.
For example, Jin \emph{et al.} \cite{jin2023fedmlhe} designed \emph{FedML-HE}, which utilizes the union of clients' sensitive model parameter masks for encryption.
Hu \emph{et al.} \cite{hu2024maskcrypt} proposed \emph{MaskCrypt}, which lets the clients agree on a consensus mask for selective HE with reduced overhead.
However, these methods cannot adaptively ensure the security of each client with low HE overhead
in cross-device FL.

\section{Comparison with FedML-HE}

Although FedML-HE \cite{jin2023fedmlhe} proposes a selective encryption mechanism based on parameter sensitivity, its efficiency is extremely low. FedML-HE calculates sensitivity via sequentially iterating the gradient value of each parameter and computing its partial derivative with respect to each data label, as formulated in
$$
J_m(y_k) = \frac{\partial}{\partial y_k} \left( \frac{\partial \ell(\mathbf{X}, \mathbf{y}, \mathbf{W})}{\partial w_m} \right),$$
where $\ell(\mathbf{X}, \mathbf{y}, \mathbf{W})$ denotes the loss function. Specifically, FedML-HE requires $\mathcal{O}(N^w \cdot N^b)$ complex backpropagation operations, where $N^w$ is the total number of model parameters and $N^b$ is the batch size. In contrast, SenseCrypt only requires $\mathcal{O}(N^w)$ simple multiplication operations, which are significantly less time-consuming than backpropagation.

Consequently, while FedML-HE required \textbf{more than 4 hours} to compute the sensitivity of 50{,}000 parameters on a 5-thread CPU, SenseCrypt and MaskCrypt completed it in \textbf{less than 0.1 seconds}. Even with an extreme space-for-time trade-off optimization, FedML-HE's computation time could only be reduced to 340 seconds, at the cost of consuming over \textbf{100~GB} of memory. Table~\ref{tab:sensitivity_time} summarizes the sensitivity computation time for a single evaluation over 200 samples on different models. Therefore, FedML-HE is unsuitable for selective encryption in heterogeneous device scenarios due to its prohibitive computational overhead.

\begin{table}[h]
\centering
\begin{tabular}{l l l r}
\toprule
\textbf{Model} & \textbf{Params} & \textbf{Dataset} & \textbf{Time (s)} \\
\midrule
MLP     &  50,890    & MNIST    & 17214.75 \\
AlexNet & 124,534  & CIFAR-10 & 42823.79 \\
\bottomrule
\end{tabular}
\caption{Sensitivity Computation Time of FedML-HE}
\label{tab:sensitivity_time}
\end{table}

\section{Data Reconstruction Example}

\begin{figure}[h!]
\centering
    \includegraphics[width = \linewidth]{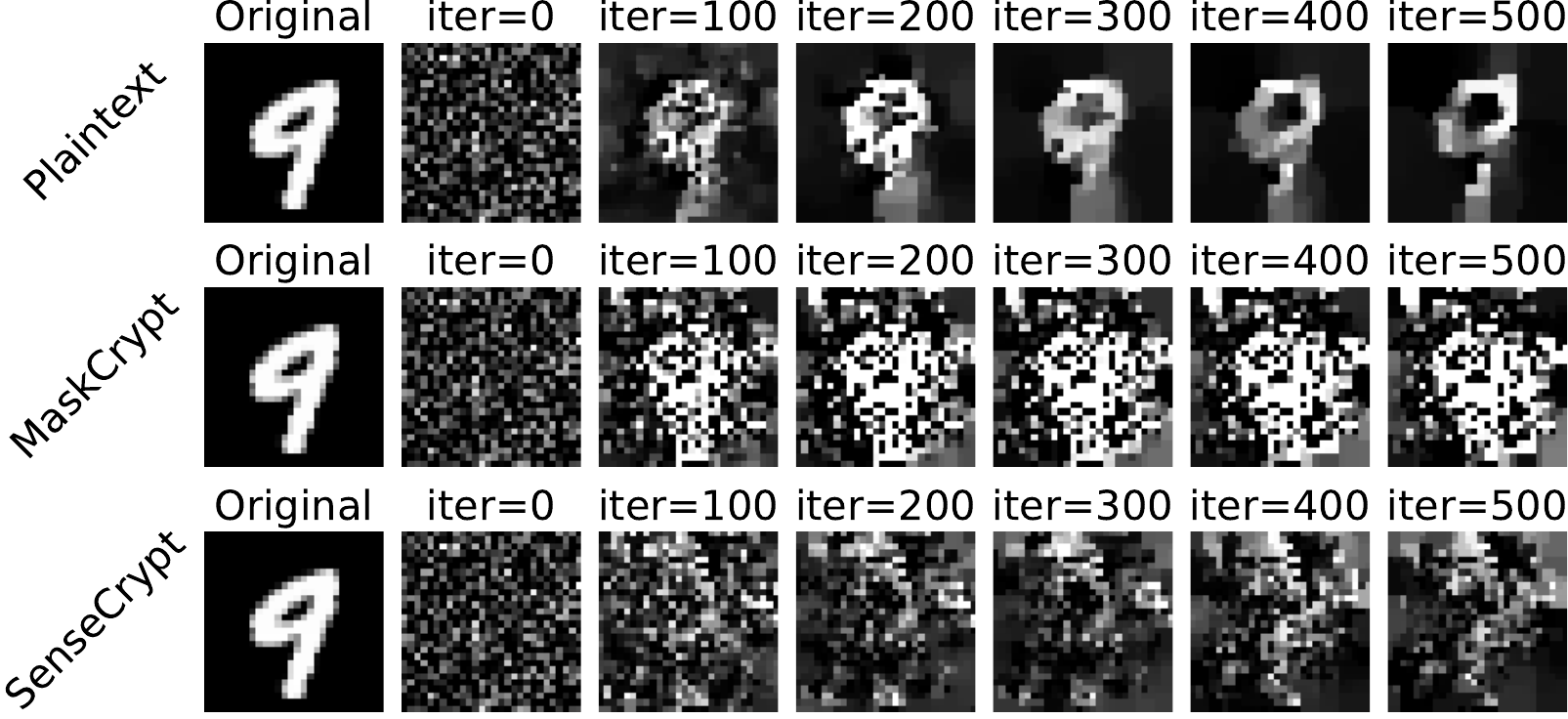}
  \caption{Comparison of data reconstruction results.}
  \label{eval_attack_illustrate}
\end{figure}

Figure \ref{eval_attack_illustrate} illustrates a \texttt{MNIST} data sample reconstructed from batch-averaged gradients via iLRG from Round 4 in Fig. 8 of the main text.
Compared with \emph{MaskCrypt}, the reconstructed data in \emph{SenseCrypt} exposes much less useful information to attackers.

\end{document}